\documentclass[letterpaper,12pt]{JHEP3}

\usepackage{bm,amsmath,amssymb,mathrsfs,slashed,fleqn}
\usepackage{epsfig,graphics,graphicx,subfigure}
\usepackage{cite}

\long\def\comment#1{ }
\newcommand{\eqnum}[1]{Eq.~\eqref{#1}}

\newcommand{\beq}{\begin{eqnarray}}
\newcommand{\eeq}{\end{eqnarray}}
\newcommand{\nn}{\nonumber\\}

\newcommand{\dif}{{\rm d}}
\newcommand{\dz}{\dif z}
\newcommand{\rmd}{{\rm d}}
\newcommand{\rme}{{\rm e}}
\newcommand{\rmi}{{\rm i}}

\newcommand{\rmH}{{\rm H}}
\newcommand{\rmI}{{\rm I}}

\newcommand{\rmK}{{\rm K}}

\newcommand{\EA}{\mcal{E}_{\rm A}}
\newcommand{\EB}{\mcal{E}_{\rm B}}

\newcommand{\del}{\partial}

\newcommand{\order}[1]{\mcal{O}{(#1)}}
\newcommand{\mcal}{\mathcal}

\newcommand{\tp}{\acute{t}}
\newcommand{\xp}{\acute{x}}
\newcommand{\zp}{\acute{z}}
\newcommand{\rp}{\acute{r}}

\newcommand{\bxT}{\bm{x}_{\perp}}
\newcommand{\br}{\bm{r}}
\newcommand{\brp}{\acute{\bm{r}}}
\newcommand{\bq}{\bm{q}}

\let\Oldcdot\cdot
\renewcommand{\cdot}{\Oldcdot}

\usepackage{accents}
\makeatletter
\def\cc@accentPhantom#1#2{%
\begingroup
\cc@code=#1\relax
\let\math@bgroup\@empty
\def\math@egroup##1{\cc@setchar##1}%
\cc@palette\cc@@accentPhantom{#2}%
\cc@nuc
\endgroup%
}
\def\cc@@accentPhantom#1#2{%
\let\cc@style=#1%
\cc@fetch{#2}%
\mathaccent\cc@code{%
\ifcc@more#2\else
\cc@phant{#2}%
\gdef\cc@nuc{}%
\fi
\kern\cc@skew%
}%
\kern-\cc@skew
\kern\cc@wd%
}
\gdef\dacute#1{%
\rlap{$\mkern-2.0mu\protect\cc@accentPhantom{"7013}{#1}$}%
\rlap{$\mkern2.0mu\protect\cc@accentPhantom{"7013}{#1}$}%
#1%
}
\makeatother

\title{\Large Aspects of the UV/IR correspondence :\\
energy broadening and string fluctuations}

\author{Y.~Hatta$^a$, E.~Iancu$^{b,c}$, A.H.~Mueller$^d$ and
D.N.~Triantafyllopoulos$^e$\\
\!\!$^a$Graduate School of Pure and Applied Sciences,
University of Tsukuba,
Tsukuba, Ibaraki 305-8571, Japan\\
\!\!$^b$CERN, Theory Division, CH-1211 Geneva, Switzerland\\
\!\!$^c$Institut de Physique Th\'{e}orique de Saclay,
F-91191 Gif-sur-Yvette, France\\
\!\!$^d$Department of Physics, Columbia University, New York,
NY 10027, U.S.A.\\
\!\!$^e$ECT*, Strada delle Tabarelle 286, I-38123 Villazzano (TN), Italy\\
E-mail: \email{hatta@het.ph.tsukuba.ac.jp,
edmond.iancu@cea.fr, amh@phys.columbia.edu, trianta@ect.it}}

\abstract{
We show that a source which radiates in the
vacuum of the strongly coupled ${\mathcal N}=4$ SYM theory produces an
energy distribution which, in the supergravity approximation,
has the same space--time pattern as the corresponding classical
distribution: the radiation propagates at the speed of light without
broadening. We illustrate this on the basis of several examples:
a small perturbation propagating down a steady string,
a massless particle falling into AdS$_5$, and the decay of a
time--like wave--packet. A similar observation was made in
{\sf Phys. Rev. D81 (2010) 126001}
for the case of a rotating string.
In all these cases, the absence of broadening is
related to the fact that the energy backreaction on the boundary
arises exclusively from the bulk perturbation at, or
near, the boundary. This is so since bulk sources which propagate in
AdS$_5$ at the speed of light do not generate any energy on the boundary.
We interpret these features as an artifact of the supergravity
approximation, which fails to encode quantum mechanical
fluctuations that should be present even in the strong coupling limit.
We argue that such fluctuations should enter the dual string theory as
longitudinal string fluctuations, which are not suppressed at strong
coupling. We heuristically estimate the effects of such fluctuations and
argue that they restore the broadening of the radiation,
in agreement with expectations from
both quantum mechanics and the ultraviolet/infrared correspondence.
}
\preprint{}

\begin{document}
\section{Introduction}

Understanding gauge theory dynamics beyond perturbation theory, and in
particular at strong coupling, represents one of the major desiderata of
theoretical physics, with ramifications going from high--energy to
condensed matter physics. Recent years have seen some important progress
in that sense thanks to the theoretical breakthrough known as the {\em
AdS/CFT correspondence} (or, with a somewhat more general sense, the {\em
gauge/string duality})
\cite{Maldacena:1997re,Gubser:1998bc,Witten:1998qj}, which allows one to
study the strong coupling regime of some special, highly symmetric, gauge
theories via weak coupling techniques in a `dual' string theory. Further
efforts in that direction have been triggered by the experimental
observation of strong coupling aspects in the dynamics of the {\em
quark--gluon plasma} --- the high--temperature, deconfined, phase of QCD,
which is produced in the intermediate stages of a heavy ion collision at
RHIC and, more recently, at the LHC. This plasma corresponds to a
physical regime where QCD itself is not so far away from its conformal
`cousins', so like the ${\mathcal N}=4$ supersymmetric Yang--Mills (SYM),
to which the AdS/CFT correspondence has been widely tested. This
observation stimulated applications of AdS/CFT to time--dependent
phenomena (see, e.g., the review papers
\cite{Son:2007vk,Iancu:2008sp,Gubser:2009sn} and refs. therein), among
which the problem of the {\em radiation} at strong coupling
\cite{Mikhailov:2003er,Herzog:2006gh,Gubser:2006bz,CasalderreySolana:2006rq,Gubser:2006nz,CasalderreySolana:2007qw,Gubser:2007xz,Chesler:2007an,Hofman:2008ar,HIM3,Gubser:2008as,Chesler:2008wd,Dominguez:2008vd,Hatta:2008st,Fadafan:2008bq,Athanasiou:2010pv,Arnold:2010ir}.

In particular, some of these studies addressed the {\em space--time
distribution} of the radiated energy, whose calculation in AdS/CFT is
quite complex, as it requires solving a gravitational backreaction
problem (see below). Motivated by the phenomenology of the quark--gluon
plasma, the first such studies were concerned with the
finite--temperature problem
--- e.g., the medium--induced radiation by a heavy quark propagating through
a strongly coupled plasma
\cite{Gubser:2007xz,Chesler:2007an,Gubser:2008as,Chesler:2008wd,Dominguez:2008vd}.
More recently, similar problems have been addressed also in the context
of the vacuum of the ${\mathcal N}=4$ SYM theory
\cite{Hofman:2008ar,HIM3,Hatta:2008st,Athanasiou:2010pv}. One of these
studies --- that of the synchrotron radiation in
Ref.~\cite{Athanasiou:2010pv} --- produced some very interesting and
intriguing results, which in particular triggered our present analysis.

Before we describe these results and our present work, let us recall that
all the AdS/CFT calculations aforementioned have been performed in the
{\em supergravity approximation} (SUGRA), which is the semiclassical
limit of the string theory in which both string loops and internal string
excitations are neglected. This approximation is generally assumed to
faithfully describe the strong `t Hooft coupling limit of the ${\mathcal
N}=4$ SYM theory, that is, the limit $\lambda=g^2N_c\to\infty$ with fixed
$g\ll 1$ ($g$ is the Yang--Mills coupling and $N_c$ the number of
colors). However, in this work we shall give arguments suggesting that
this is not always the case: SUGRA seems unable to capture the detailed
space--time distribution of the radiation emitted in the vacuum and in
the strong coupling limit. This is so since, as we shall demonstrate,
this distribution is affected by {\em longitudinal string fluctuations}
which are {\em not} suppressed in the strong coupling limit.

Our arguments in that sense will be constructed in two steps, with
different degrees of rigor: \texttt{(1)} a {\em supergravity calculation}
of the energy backreaction in Sects.~\ref{sect-classical} and
\ref{sect-falling}, which is very explicit and in some cases even exact,
and \texttt{(2)} a calculation of the {\em string fluctuations} in
Sect.~\ref{sect-flucts}, which is semi--heuristic (in the sense of using
string quantization in flat space--time), but which we believe to capture
the salient features of the actual situation in curved space--time. In
what follows, we shall describe these two steps in more detail.

The first step involves explicit calculations of the energy distribution
produced in the supergravity approximation by various types of sources
radiating in the vacuum of ${\mathcal N}=4$ SYM. One purpose of this
analysis is to demonstrate that the main result in
Ref.~\cite{Athanasiou:2010pv} is in fact {\em generic} (and not specific,
say, to the special geometry of the rotating quark). Namely, the
supergravity prediction for the radiation emitted by an arbitrary source
exhibits {\em the same space--time pattern} as the corresponding {\em
classical} result, without any trace of broadening. By `broadening' we
mean off--shell effects, which would be natural in a quantum field theory
(and even more so at strong coupling !) and would yield components of the
radiation which propagate slower than light, thus leading to a spreading
in the radiation emitted by a source which is localized in space and
time. But our calculations show that, in the SUGRA limit, the radiation
always propagates at the speed of light, so like the solution to the
classical Maxwell equations. For instance, a point--like source at $r=0$
which is a pulse in time with duration $\sigma$ generates a thin
spherical shell of energy, which is localized at $r=t$ with a width equal
to $\sigma$.

One reason why this lack of broadening looks so surprising is because, at
a first sight at least, it seems to contradict the {\em
ultraviolet/infrared (UV/IR) correspondence}
\cite{Susskind:1998dq,Peet:1998wn}, which is one of the pillars of the
gauge/string duality. In order to explain this puzzle, let us first
recall some elements of the AdS/CFT formalism and thus fix some
notations.


 The five--dimensional (5D) Anti de Sitter space--time
AdS$_5$ where the string theory is formulated can be viewed as a product
between the 4D Minkowski space--time (the `boundary' of AdS$_5$) times a
`radial'\footnote{Following standard conventions, we shall use the word
`radial' in relation with both the 5th dimension $z$, and the physical
radius $r=|\br|$ in 3 dimensions. The precise meaning should be clear
from the context.} (or `fifth') dimension, that we shall denote as $z$,
in conventions where $0\le z<\infty$ and the Minkowski boundary lies at
$z=0$ (see \eqnum{metric} for the precise form of our metric). Roughly
speaking, this fifth dimension acts as a reservoir of quantum
fluctuations for the dual gauge theory. The UV/IR correspondence is a
more precise version of this statement. This is a unique tool allowing
for the physical interpretation of the results of the string theory back
to the original gauge theory.

Specifically, the UV/IR correspondence relates the typical
energy/momentum scales (or, by the uncertainty principle, the space--time
scales) of the physical phenomenon on the `boundary' to the radial ($z$)
position of the dual string excitation in the `bulk'. For real--time
phenomena like radiation, there are at least two types of scales which
can be encoded in this way: \texttt{(i)} the {\em overall size} $R$ of
the phenomenon on the boundary\footnote{Here $R$ refers to a frame where
the center of mass of the energy distribution is at rest; in other
frames, one must take into account the Lorentz contraction (see e.g. the
discussion in Sect.~\ref{Constant}).}, which with our present conventions
is proportional to the radial penetration $z$ of the dual excitation in
the bulk, and \texttt{(ii)} the {\em space--time virtuality} $\Delta s^2$
on the boundary (say, the difference $t^2-r^2$ in the case of the
spherical shell of radiation alluded to above), which is proportional to
the position $z$ of the string source in the bulk {\em at the time of
emission}. The last statement refers to the calculation of the energy
density on the boundary from the response of the AdS$_5$ metric to the 5D
stress tensor of the string excitation in the bulk (which for this
purpose plays the role of a `source'). This is the calculation generally
referred to as the `backreaction'.


How do these scaling arguments apply to the supergravity results for
radiation at strong coupling ? First of all, we should say that they work
fine for the radiation emitted in a {\em medium}
\cite{Gubser:2007xz,Chesler:2007an,Chesler:2008wd,Dominguez:2008vd}. For
instance, a heavy quark moving at constant speed in the  ${\mathcal N}=4$
SYM plasma at temperature $T$ is dual to the `trailing string' --- a
Nambu--Goto string attached to the quark and pulled by the latter through
the AdS$_5$ black hole geometry representing the plasma. Each bit of this
string, which in the 5th direction extends from the Minkowski boundary
down to the black hole horizon at $z\sim 1/T$, yields a contribution to
the energy on the boundary with a transverse width proportional to the
location $z$ of that bit
\cite{Gubser:2007xz,Chesler:2007an,Dominguez:2008vd}. In particular, when
$z$ (and hence the transverse width) becomes of order $1/T$, this is a
sign that the respective component of the radiation has thermalized (and
then propagates as a hydrodynamical wave in the medium
\cite{Chesler:2007an}).

Returning to the problem of the radiation in the {\em vacuum}, we note
that the first `UV/IR correlation' above --- that between the overall
size and the radial penetration of the dual excitation in the bulk --- is
indeed respected, as is easy to see for a localized excitation. For
instance, for the `pulse on the boundary' alluded to above (more
precisely, a time--like wave--packet which can represent the virtual
photon produced in electron--positron annihilation), we shall find that
the dual bulk perturbation propagates like a massless particle falling
into AdS$_5$ (a wave--packet centered at $z=t$). This is in agreement
with the fact that, on the boundary, the energy density expands as a
spherical shell at $r=t$ (see the discussion in Sects.~\ref{Falling} and
\ref{TLWP}). A similar situation holds when the physical source is a
heavy quark whose dual object in the bulk is a Nambu--Goto string. Then,
the bulk perturbation which propagates at the 5D speed of light is a {\em
bit of energy} flowing down the string. (The calculation of this 5D
velocity requires a careful separation of the radiative part from the
Coulomb part of the energy, which are both encoded in the Nambu--Goto
string; see Sect.~\ref{Speed} and the subsequent publication \cite{wip}
for more details.) This is in agreement with the fact that the radiation
produced via backreaction on the boundary expands outward at the speed of
light, as found in \cite{Athanasiou:2010pv} for the rotating string and
as we shall see in Sect.~\ref{SmallPulse} below for a small but generic
string perturbation.

What about the other `UV/IR correlation', that between the {\em
space--time virtuality} of the radiation on the boundary and the position
$z$ of its source in the bulk ? At first sight, this seems to be violated
by the SUGRA results that we are aware of: the radiation propagates in
the vacuum without broadening, so like {\em massless, on--shell, quanta},
in spite of the fact that there are sources in the bulk at arbitrary
large distances $z$. Such a violation would be very strange though: as we
shall argue in Sect.~\ref{SmallPulse}, the correlation between the
broadening of the distribution on the boundary and the position of the
source in the bulk is simply a consequence of causality. (Such a
correlation is for instance manifest in the calculation of the `glueball'
density ${\rm Tr}\,F_{\mu\nu}^2$ produced on the boundary by a small
perturbation of a static string \cite{Callan:1999ki}.) And indeed a
closer inspection of the calculations in Sects.~\ref{sect-classical} and
\ref{sect-falling} and also in Ref.~\cite{Athanasiou:2010pv} reveals that
this apparent contradiction is solved by the fact that, in all these
calculations, {\em the whole backreaction on the boundary} (in terms of
radiation) {\em comes from sources in the bulk which are located close to
the boundary}, within a distance in $z$ set by the width $\sigma$ of the
physical perturbation. In particular, in the limit where the physical
source has zero width, the backreaction on the boundary comes exclusively
from the limit $z\to 0$ of the perturbation in the bulk\footnote{In fact,
even in more general cases where the physical source is delocalized on
the boundary (like the rotating quark in Ref.~\cite{Athanasiou:2010pv})
it appears that, via appropriate changes of variables in the convolutions
expressing the backreaction, one can obtain the whole result for the
energy density as an endpoint contribution, coming from the boundary
endpoint ($z=0$) of the integral over $z$ \cite{Athanasiou:2010pv,wip}.}.

But whereas it solves a potential conflict with the UV/IR correspondence,
this last observation introduces another puzzle: why should bulk sources
which lie far away from the boundary have no reflection in terms of
energy density in the physical gauge theory~? Our investigations will
allow us to elucidate the mathematical origin of this mysterious property
in the context of the SUGRA calculations: it arises from the fact that
bulk sources which propagate in AdS$_5$ at the speed of light do not
generate any energy backreaction on the boundary\footnote{The importance
of the bulk propagation at the speed of light for the problem of the
synchrotron radiation in AdS/CFT has been also recognized in the recent
paper \cite{Hubeny:2010rd}. There it is shown that each energy bit
flowing down the rotating string generates a backreaction in the form of
a gravitational shock--wave and that, by superposing such shock--waves
for all the bits along the string, one obtains a pattern for the energy
density on the boundary which is similar to that found in
\cite{Athanasiou:2010pv}. However, it was not realized there that each of
these shock--waves is in fact generated via backreaction from $z\approx
0$ (that is, from the very early stages of the trajectory of each energy
bit) and that this is the reason why the resulting energy distribution
shows no more broadening than the classical one.}. This property is quite
intuitive when the source is moving {\em parallel to the boundary}, that
is, it stays at a fixed value of the 5th coordinate $z$~: then, the
backreaction is a gravitational shock--wave \cite{Gubser:2008pc} which,
because of Lorentz time dilation, is fully generated at very early times
$t\to-\infty$ (see Sect.~\ref{Constant} below). In that case, there is
naturally no energy backreaction from any {\em finite} value of $t$.
Similarly, for the case of a massless particle falling into AdS$_5$ at
the speed of light (say, along the 5th dimension), we shall find that
there is no energy backreaction except from the starting point of the
trajectory at $t=z=0$. The corresponding backreaction is a gravitational
shock--wave with zero width (since fully generated at $z=0$) and whose
intersection with the boundary is a spherical shell moving outward with
$r=t$.

For motion at the speed of light which is oriented towards the interior
of AdS$_5$, one can {\em formally} interpret the absence of energy
backreaction as a kind of `Lorentz time dilation', but only after
performing a special change of coordinates which mixes the fifth
dimension with one of the spatial dimensions (see Sect.~\ref{sect-LC}).
In these new coordinates, the propagation of the radiation without
broadening appears as `Lorentz contraction'. But we do not find this
interpretation as very natural, or useful.

More generally, we believe that there is no physical foundation for this
property of the supergravity approximation, which is the lack of
broadening for radiation at strong coupling. Indeed, this seems difficult
to reconcile with general expectations about the dynamics in an
interacting quantum field theory, like the importance of off--shell
effects generated by virtual quantum fluctuations. In particular, this
property is inconsistent with a physical picture for parton branching at
strong coupling \cite{HIM3}, which is supported by other AdS/CFT
calculations, like those concerning the anomalous dimensions for the
leading--twist operators
\cite{Gubser:2002tv,Kotikov:2004er,Brower:2006ea,Hatta:2008tn}, the
structure functions for deep inelastic scattering
\cite{Polchinski:2002jw,HIM1,HIM2,Avsar:2009xf}, or the angular
distribution of the energy density produced via the decay of a time--like
wave--packet \cite{Hofman:2008ar}. Note however that none of these
previous calculations has investigated the detailed space--time
distribution of the energy density, so they could not encounter the
difficulties that we are concerned with here.

Motivated by such considerations which make us feel uncomfortable with
the supergravity prediction for the energy distribution, we shall proceed
in Sect.~\ref{sect-flucts} to a study of the stability of this
approximation with respect to string fluctuations. Such a study is
necessarily heuristic since the quantization of the string fluctuations
in curved space--time is an unsolved problem. Here, we shall follow a
pragmatic approach proposed by Hofman and Maldacena \cite{Hofman:2008ar},
which consists in using string quantization in flat space and the
light--cone (LC) gauge together with a special change of coordinates,
introduced in \cite{Cornalba:2007fs}, which mixes the fifth coordinate
$z$ with one of the spatial coordinates (see Sect.~\ref{sect-LC}). Hofman
and Maldacena used this strategy to study the angular distribution of the
energy produced on the boundary by a massless particle falling in AdS$_5$
(the bulk excitation dual to a time--like photon). They computed only the
energy density integrated over the spatial radius $r$, so they could not
notice that the energy is actually localized on a thin shell at $r=t$.
For that particular set--up, they have shown that the energy distribution
in the SUGRA approximation is isotropic {\em event by event} and that the
string corrections to it --- which involve the {\em transverse} string
fluctuations --- are suppressed, as expected, in the strong coupling
limit.

Here, however, we shall be concerned with the {\em radial} ($r=|\br|$)
distribution of the energy density, which in the coordinates of
Ref.~\cite{Cornalba:2007fs} receives corrections from the {\em
longitudinal} string fluctuations. The latter are not independent degrees
of freedom, rather they are related to the transverse fluctuations via
the constraints of the LC gauge, but this relation is such that the
longitudinal fluctuations are {\em not} suppressed as $\lambda\to\infty$.
This last property is well known (it is manifest in textbook treatments
of string quantization in flat space
\cite{Polchinski:1998rq,Zwiebach:2004tj}), yet we are not aware of any
application of it in the context of AdS/CFT. This is most likely so
because the longitudinal degrees of freedom do not directly participate
in standard string problems like the scattering between two strings.

Following the strategy in \cite{Hofman:2008ar}, in Sect.~\ref{Longit} we
shall show that the radial distribution of the energy density receive
sizeable corrections from the longitudinal string fluctuations. These
corrections are independent of $\lambda$ --- at least up to issues of
ultraviolet divergences, that we shall shortly comment on. Hence they are
a part of the leading order result in the strong coupling limit and they
act, as expected, towards spreading the distribution in $r-t$. Moreover
this spreading is such that the UV/IR correspondence is respected: a
contribution to the energy density from string fluctuations at $z$ yields
a spreading $t-r\sim z$.

Our present analysis cannot be seen as definitive, first, because of the
lack of rigor in the string quantization prescription and, second,
because the analysis in Sect.~\ref{Longit} is plagued with ultraviolet
divergences, as is generally the case for problems involving string
fluctuations in flat space. Polchinski and Susskind have argued that, in
AdS$_5$ the UV divergences should be cured by the warp factor
\cite{Polchinski:2001ju}. We do not know whether the arguments in
Ref.~\cite{Polchinski:2001ju} can be extended to the longitudinal
fluctuations of interest to us here. But independently of such issues,
which require further clarifications, we believe that our results provide
solid evidence in favor of the failure of the supergravity approximation
for the observables which are sensitive to longitudinal string
fluctuations.

\section{Classical strings}
\label{sect-classical}

In this section, we shall present our first example of a supergravity
calculation of radiation at strong coupling, whose result shows the
remarkable feature outlined in the Introduction: the radiated energy
propagates without broadening, in spite of having bulk sources at
arbitrarily large distances in the 5th dimension.

Specifically, in Sect.~\ref{SmallPulse} we shall compute the energy
density radiated by a small perturbation (`pulse') propagating along a
steady string. In physical terms, the string describes the wavefunction
of a heavy quark and the pulse represents its response to an external
perturbation acting for a short lapse of time. The condition that the
string perturbation remain small is equivalent to the non--relativistic
approximation for the motion of the heavy quark. We shall find that the
whole contribution to radiation (as obtained after subtracting the
Coulomb piece of the energy) is generated via backreaction from the
string endpoint at the boundary {\em alone}. Via the UV/IR
correspondence, this property is correlated with the lack of broadening
alluded to above.

But before turning to the string perturbation, we shall first review, in
Sect.~\ref{Constant}, the calculation of the energy backreaction for an
unperturbed string moving at constant speed. Besides allowing us to
introduce the general formalism in a simpler setup, this problem is
interesting also in that it offers a conceptually simple example where
the UV/IR correspondence works as expected --- in relation with the
Coulomb energy of the heavy quark. Moreover, in this context we shall for
the first time observe a property which will play an important role later
on: a bulk source which propagates at the 5D speed of light does not
generate any energy on the Minkowski boundary.

Finally in Sect.~\ref{Speed} we shall argue that this last property is
responsible for the lack of backreaction from string points away from the
boundary, for both the small string perturbation studied in
Sect.~\ref{SmallPulse} and the rotating string problem discussed in
Ref.~\cite{Athanasiou:2010pv}.

\subsection{Heavy quark with constant velocity}
\label{Constant}

Consider an on--shell heavy quark moving with constant velocity
$\upsilon<1$ along the $x\equiv x^1$ axis within the vacuum of the
${\mathcal N}=4$ SYM theory. In the dual string theory, and with a
convenient choice of coordinates in AdS$_5$, this is described as a
Nambu--Goto string attached to a D7--brane hanging vertically along the
radial direction of AdS$_5$ and moving at constant speed
$\upsilon^M=\upsilon\,\delta^{M1}$. ($M=0,1,2,3,5$ are vector indices for
AdS$_5$ : the first four indices refer to the Minkowski components, and
the 5th one to the radial direction.) Specifically, using the so--called
Poincar\'e coordinates,
 \beq\label{metric}
\rmd s^2\,\equiv\,G_{MN}\rmd x^M \rmd x^N
 \,= \,
\frac{L^2}{z^2} \,\big[ -\rmd t^2 + \rmd \br^2 + \rmd z^2\big],
 \eeq
together with the temporal--gauge parametrization $X^M(\tau,\sigma)\equiv
X^M(t,z)$ for the string world--sheet, the string embedding function
reads $X^M=(t, \upsilon t, 0, 0, z)$. In \eqnum{metric},
$\br=(x^1,x^2,x^3)$ and we shall often use the notations $x\equiv x^1$
and $\bxT \equiv (x^2, x^3)$. The variable $z$ along the string is
restricted to $z\ge z_m$ where $z_m= {\sqrt{\lambda}}/(2\pi m_q)$ is the
radial position of the D7--brane and $m_q$ is the mass of the heavy
quark. In what follows we shall assume the quark to be heavy enough for
$z_m$ to be much smaller than all the other scales in the problem (e.g.,
the space--time locations where we measure the energy density), and we
shall often set $z_m=0$ in explicit calculations.

The moving quark generates (color) electric and magnetic fields, which
are simply the boosted versions of the Coulomb field produced by the
quark in its rest frame. We would like to compute the energy density $
\mcal{E}\equiv \langle T_{00}\rangle$ which is stored in these fields.
Within the supergravity context, this is obtained from the `backreaction'
of the associated Nambu--Goto string on the Minkowski boundary
\cite{deHaro:2000xn,Skenderis:2002wp}. The respective construction is in
fact more general than the specific quark/string problem at hand
--- it applies whenever we have a `source' of energy and momentum in the bulk,
with 5D stress tensor $t_{MN}$, which is dual to some physical
perturbation, or bound state, in the boundary gauge theory.

In what follows, we shall exhibit the general formul\ae{} expressing the
energy backreaction for the type of problems of interest to us here ---
namely, for bulk sources which are dual to small perturbations of the
${\mathcal N}=4$ SYM vacuum. (This is in particular the case for the
heavy quark problem under consideration.) By `small perturbations', we
more precisely mean sources $t_{MN}$ which  in the limit $N_c\to\infty$
scale like $N_c^0=1$, so the associated change $h_{MN}\equiv \delta
G_{MN}$ in the bulk metric is a small effect of $\order{1/N_c^2}$. To
compute this effect, it is enough to solve the linearized version of the
5D Einstein equations. The gauge theory stress tensor $\langle
T_{\mu\nu}\rangle $ is finally inferred from a study of the behaviour of
the metric perturbation $h_{MN}$ near the Minkowski boundary at $z=0$. By
working in a gauge where $h_{5M}=0$ and denoting
 \beq
 H_{\mu\nu} \equiv \frac{z^2}{L^2}\,h_{\mu\nu} =
 H^{(3)}_{\mu\nu} z^3 + H^{(4)}_{\mu\nu} z^4 + \order{z^5},
 \eeq
one can compute the gauge theory expectation value of the stress tensor
from the coefficient of $z^4$ in the above small--$z$ expansion :
 \beq
  T_{\mu\nu}  = \frac{2 L^3}{\kappa_{5}^2}\,H^{(4)}_{\mu\nu} =
 \frac{N_c^2}{2 \pi^2}\,H^{(4)}_{\mu\nu}.
 \eeq
Although it amounts to solving linearized equations of motion, the
calculation of $H_{MN}$ is quite involved, because of the many coupled
equations for the (generally) fifteen independent components. However,
this calculation can be more economically organized by recognizing that
the fifteen `independent' components of $H_{MN}$ depend upon only five
gauge--invariant degrees of freedom, in agreement with the number of
physical degrees of freedom expected for the 4D stress tensor
$T_{\mu\nu}$. These physical degrees of freedom (linear combinations of
$H_{MN}$ and its derivatives) can be chosen in (arbitrarily) many ways,
and here we shall follow the procedure in \cite{Athanasiou:2010pv}, from
which we shall simply quote the relevant results (see
Refs.~\cite{Chesler:2007an,Chesler:2007sv} for other such choices).

Namely, specializing to $\mcal{E} \equiv T_{00}$ and using Eq.~(3.62) in
Ref.~\cite{Athanasiou:2010pv}, one finally arrives at the following
expression for the energy density at a space--time point $x^\mu=(t,\br)$
on the boundary:
 \beq
\mcal{E}(t,\br) \,=\,\EA(t,\br)\,+\,\EB(t,\br)
 \eeq
where
 \beq\label{EA}
 \hspace*{-1.3cm}\EA &=& \frac{2 L^3}{\pi}
 \int \frac{\dif^4\rp \,\dif z}{z^2}\, \Theta(t-\tp) \delta''(\mcal{W})
 \left[z (2 t_{00} - t_{55}) - (t-\tp) t_{05} + (x - \xp)^i t_{i5}
 \right],
 \\ \label{EB}
 \hspace*{-1.3cm}\EB &=& \frac{2 L^3}{3 \pi}
 \int \frac{\dif^4\rp \,\dif z}{z}\, \Theta(t-\tp) \delta'''(\mcal{W})
 \left[|\br - \brp|^2 (2 t_{00} -2 t_{55} + t_{ii}) -
 3 (x - \xp)^i (x - \xp)^j t_{ij} \right].
 \eeq
In these equations, the bulk point with coordinates $(\tp,\brp,z)$ is the
source point from which originates the perturbation and $\dif^4\rp\equiv
\dif\tp \,\dif^3\brp$. (Throughout this paper, we shall systematically
use the symbol acute for the spatial coordinates of a point belonging to
a source in the bulk.) Furthermore, the quantity
 \beq\label{W}
 \mcal{W} = -(t-\tp)^2 + (\br - \brp)^2 +z^2\,.
 \eeq
is proportional to the 5D invariant distance between that source point in
the bulk and the point of measurement on the boundary. The
$\delta$--function $\delta(\mcal{W})$, which enters via the (retarded)
bulk--to--boundary propagator \cite{Athanasiou:2010pv}, shows that the
metric perturbation propagates throughout AdS$_5$ at the 5D speed of
light (which is equal to one in our present conventions), as expected for
classical, massless fields in AdS$_5$ --- in this case, gravitational
waves. The presence of derivatives of the $\delta$--function may seem
peculiar, but it is a generic feature of retarded propagators in AdS, as
we shall show in Appendix \ref{appProp} where we construct the respective
scalar propagator. The derivatives of $\delta(\mcal{W})$ in
Eqs.~\eqref{EA} and \eqref{EB} can be taken as derivatives w.r.t. one of
the external variables appearing in $\mcal{W}$ and subsequently pulled
out of the integrand, but caution is needed since there might be explicit
dependence on that variable in the integrand. In such cases it is better
to write
 \beq
 \delta^{(n)}(\mcal{W}) = \lim_{\epsilon \to 0}
 \del_{\epsilon}^n \delta(\mcal{W}+\epsilon).
 \eeq

As already emphasized, Eqs.~\eqref{EA} and \eqref{EB} hold for an
arbitrary bulk source with (parametrically small) stress tensor $t_{MN}$.
When this source is a string --- the case of interest in this section
---, there are some simplifications due to the fact that, for a string,
$t_{MN}$ has support only on the 2D string world--sheet. Using the
parametrization $X^M=(t, \br_s(t,z), z)$ for the latter, one has $t_{MN}
(\tp,\brp,z)\propto \delta^{(3)}(\brp-\br_s(\tp,z))$ and then the spatial
integrations in Eqs.~\eqref{EA} and \eqref{EB} are trivially done.

More explicitly, for a generic string profile $X^M=(t, \br_s(t,z), z)$,
one has
\begin{equation}\label{tMN} t^{MN}(t, \br, z)\, =\,
-\frac{T_0}{\sqrt{-G}}\sqrt{-g}\,g^{ab}\,
\partial_a X^M \,\partial_b X^N\,
\delta^{(3)}(\bm r-\bm r_s)\,.
\end{equation}
where $T_0 = \sqrt{\lambda}/{2 \pi L^2}$ is the string tension, $G_{MN}$
is the (unperturbed) AdS$_5$ metric from \eqnum{metric}, and $g_{ab}$,
with $a,b=t,z$, is the induced metric on the string world--sheet:
 \beq
 g_{ab}\,=\,G_{MN}\del_a X^M \del_b X^N\,,\quad
 -g\,=\,\big(\del_t X\cdot \del_z X\big)^2 -
 \big(\del_t X\big)^2\big(\del_z X\big)^2\,.
 \eeq
In what follows, we shall often use a dot (prime) to denote a derivative
w.r.t. $t$ ($z$).

Armed with such general formul\ae, we now return to the case of a heavy
quark at constant velocity, for which we have $X^M=(t, \upsilon t, 0, 0,
z)$. The respective 5D bulk stress energy tensor is easily obtained as
 \beq
 t_{MN} = \frac{\sqrt{\lambda}}{2 \pi}\,\frac{\gamma z}{L^3}\,
 \delta^{(2)}(\bxT)\,\delta(x-\upsilon t)
 \begin{pmatrix}
 1 & -\upsilon & 0 \\
 -\upsilon & \upsilon^2 & 0\\
 0 & 0 & \upsilon^2-1
 \end{pmatrix},
 \eeq
where, with increasing order, the elements correspond to the $t$, $x$ and
$z$-components, while all the elements which involve at least one
transverse component are identically zero and are not shown. $\gamma=
1/\sqrt{1- \upsilon^2}$ is the Lorentz contraction factor. Using
\eqnum{EA} we find
 \beq\label{EAunif}
 \EA = \frac{\sqrt{\lambda}}{\pi^2}\,\gamma (3-\upsilon^2)
 \del^2_{\bxT^2}\int \dz\, \dif \tp \,
 \delta(\mcal{W})\,,
 \eeq
where for the situation at hand
 \beq\label{Wunif}
 \mcal{W} = -(t-\tp)^2 + (x- \upsilon \tp\,)^2+\bxT^2 +z^2\,.
 \eeq
Solving the constraint $\mcal{W}=0$ and taking into account only the
retarded solution, we write
 \beq\label{tpHQ}
 \delta(\mcal{W}) = \frac{\gamma \delta[\tp - \gamma^2 (t - \upsilon x)
  + \gamma
 \sqrt{\bxT^2 + z^2+ \gamma^2 (x - \upsilon t)^2}]}
 {2\sqrt{\bxT^2 + z^2+ \gamma^2 (x - \upsilon t)^2}}.
 \eeq
Integrating over $\tp$ and taking the two derivatives we obtain
 \beq\label{EAHQ}
 \EA = \frac{3 \sqrt{\lambda}}{8 \pi^2}\,\gamma^2 (3-\upsilon^2)
 \int\frac{\dz}{[\bxT^2 + z^2+ \gamma^2 (x - \upsilon t)^2]^{5/2}}\,.
 \eeq
The integral over $z$ can be easily performed to yield
 \beq\label{EAHQ1}
 \EA = \frac{\sqrt{\lambda}}{4 \pi^2}\,
 \frac{\gamma^2 (3-\upsilon^2) }{[\bxT^2 + \gamma^2 (x - \upsilon t)^2]^2}
 \,.
 \eeq
Here, however, we are less interested in the final result for the energy
density on the boundary, but more in the way how this result gets built
by adding contributions coming from different points $z$ inside the bulk.
We shall return to this issue in a moment, but before that let us also
compute the second term $\EB$. Starting from \eqnum{EB} we have
 \beq
 \EB = \frac{\sqrt{\lambda}\gamma}{3 \pi^2}\,\lim_{\epsilon \to 0} \del_{\epsilon}^3 \int \dz\, \dif \tp \, \delta(\mcal{W}+ \epsilon)
 \left[(4- \upsilon^2) \bxT^2 + 4 (1-\upsilon^2) (x- \upsilon \tp)^2\right].
 \eeq
It is straightforward to integrate over $\tp$ using the
$\delta$-function, differentiate three times w.r.t. $\epsilon$, set
$\epsilon=0$ and finally integrate over $z$ to arrive at
 \beq\label{EBHQ}
 \EB = - \frac{\sqrt{\lambda}\gamma^2}{6 \pi^2}\,
 \frac{(4- 2\upsilon^2)[\bxT^2 + \gamma^2 (x - \upsilon t)^2] - \upsilon^2\gamma^2 (x - \upsilon t)^2}{[\bxT^2 + \gamma^2 (x - \upsilon t)^2]^3}.
 \eeq
Adding the two contributions we obtain the expected result
\cite{Gubser:2007xz,Chesler:2007an,Dominguez:2008vd}
 \beq
 \mcal{E} = \frac{\sqrt{\lambda}\gamma^2}{12 \pi^2}\,
 \frac{(1+ \upsilon^2)\bxT^2 + (x - \upsilon t)^2}
 {[\bxT^2 + \gamma^2 (x - \upsilon t)^2]^3}.
 \eeq
When $\upsilon \to 1$ this formula approaches a shock--wave, as expected
from Lorentz contraction:
 \beq\label{SW}
 \mcal{E} = \frac{\sqrt{\lambda}\gamma}{16 \pi\bxT^3}\,
 \delta(x-t).
 \eeq
This is in agreement with the shock--wave constructed in
\cite{Gubser:2008pc,Gubser:2009sx} when the latter is integrated over all
source positions $z_*$ in the bulk with a weight proportional to
$1/z_*^2$, as suggested from the expression of the static energy of a
string hanging down.

We now turn to our main interest in this calculation, which is to show
that the UV/IR correspondence is indeed satisfied in this case, in a
rather precise, quantitative way. Specifically, a bit of the string
located at radial distance $z$ from the boundary contributes
predominantly to the Coulomb energy density around the space--time points
$x^\mu=(t,x,\bxT)$ with $x_\perp \sim \gamma(x-\upsilon t) \sim z$ (here,
$x_\perp=|\bxT|$). This is simply the Lorentz boosted version of the
(perhaps more familiar) statement that, in the rest frame of the heavy
quark ($\upsilon=0$), a bit of the string at $z$ is responsible for the
backreaction at points $\br$ with $r\equiv |\br|\sim z$. This follows by
inspection of the convergence properties of the integral over $z$ in
Eq.~\eqref{EAHQ} and the corresponding one for $\EB$.

Consider \eqnum{EAHQ} for definiteness and assume that  $x_\perp \sim
\gamma(x-\upsilon t)$ and they are both very large as compared to the
lower limit $z_m$ in the integral over $z$ (which has been kept implicit
in equations like \eqref{EAunif}). We can divide the integration range
into three domains: \texttt{(i)} $z\ll x_\perp$, \texttt{(ii)} $z\simeq
x_\perp$, and \texttt{(iii)} $z\gg x_\perp$. In domain \texttt{(i)}, we
can neglect $z$ within the integrand and the ensuing integral, that is,
 \beq
I_1\approx \int_{z_m}^{x_\perp}\frac{\dz}{[\bxT^2 + \gamma^2 (x -
\upsilon t)^2]^{5/2}}\,\approx\,\frac{x_\perp}{[\bxT^2 + \gamma^2 (x -
\upsilon t)^2]^{5/2}}\,\sim\,\frac{1}{x_\perp^4}\,,
 \eeq
is dominated by its upper limit $z\simeq x_\perp$ (at least
parametrically and for relatively large values of $x_\perp$). In domain
\texttt{(iii)}, the external coordinates can be neglected and then
 \beq
I_3\approx
\int_{x_\perp}^\infty\frac{\dz}{z^5}\,\sim\,\frac{1}{x_\perp^4}\,,
 \eeq
where this time the integral is dominated by its lower limit, that is,
$z\simeq x_\perp$ once again. Thus, the integral is indeed controlled by
values $z\sim x_\perp$. This is the expected manifestation of the UV/IR
correspondence for the problem at hand. Note also the way how this
correspondence works in the presence of a Lorentz boost: the longitudinal
extent $x-\upsilon t$ of the energy generated by a bit of string at $z$
scales like $x-\upsilon t\sim z/\gamma$; this is the Lorentz--contracted
version of the corresponding extent $x_\perp\sim z$ in transverse
directions.

The second issue of interest to us here is the typical value of the
`emission' time $\tp$ which contributes to the backreaction. We would
like to show that, in the ultrarelativistic limit $v\to 1$, or
$\gamma\to\infty$, the whole contribution to the energy density --- which
in that limit takes the form of the shock--wave \eqref{SW} --- comes from
very early times $\tp\to -\infty$. The corollary of that is that a bit of
the string which propagates through AdS$_5$ at the 5D speed of
light\footnote{Note that for the uniformly moving heavy quark, the 4D
velocity of the heavy quark on the boundary coincides with the 5D
velocity of any bit of the vertical string in AdS$_5$.} does not produce
any backreaction on the boundary (since there is no contribution to the
energy density \eqref{SW} from any finite $\tp$).

To show that, we shall rely on the expression for the emission time that
can be read off \eqnum{tpHQ}, that is,
 \beq\tp \,=\, \gamma^2 (t - \upsilon x)
  - \gamma
 \sqrt{\bxT^2 + z^2+ \gamma^2 (x - \upsilon t)^2}\,.
 \eeq
Let us focus on a bit of string at a fixed value of $z$. We have just
seen that, this particular piece of string will contribute to the
backreaction at $x_\perp \,, \gamma(x-\upsilon t) \sim z$. When
$\gamma\to\infty$, the response is peaked at $x-t\sim z/\gamma\to 0$ (as
also manifest on \eqnum{SW}) and then it is convenient to write $x\equiv
t -\delta\bar x/\gamma$, where the quantity $\delta\bar x \sim z$ thus
defined remains fixed as $v\to 1$. Simple algebra using $1-v \simeq
1/2\gamma^2$ as $v\to 1$ implies
 \beq
 t-\upsilon x &=& t(1-\upsilon) + \upsilon\frac{\delta\bar x}{\gamma}
  \,\simeq\,\frac{t}{2\gamma^2} + \frac{\delta\bar x}{\gamma}\,,\nn
  x-\upsilon t &=& t(1-\upsilon) - \frac{\delta\bar x}{\gamma}
  \,\simeq\,\frac{t}{2\gamma^2} - \frac{\delta\bar x}{\gamma}\,,\eeq
and hence
 \beq\tp \,\simeq\, \frac{t}{2}  +\gamma \delta\bar x - \gamma
 \sqrt{\bxT^2 + z^2+ [(t/2\gamma) + \delta\bar x]^2}\ \to\
 \gamma \delta\bar x - \gamma
 \sqrt{\bxT^2 + z^2+ \delta\bar x^2}\,,
 \eeq
where the last limit hold as $v\to 1$ at fixed $t$. Clearly $\tp < 0$ for
any $t$ and finite value of $\gamma$, and $\tp\to -\infty$ when
$\gamma\to\infty$, as anticipated.

For the problem at hand, where the 5D velocity of any bit of string is
parallel to the Minkowski boundary, the above property is easy to
understand : this is merely a consequence of Lorentz time dilation. But
as we shall later discover, the same mathematical property --- the fact
that there is no boundary backreaction from sources which propagate
inside AdS$_5$ at the 5D speed of light --- holds whatever the direction
of motion of the sources inside the bulk and in particular for sources
falling along $z$.


\subsection{Small pulse down the string}
\label{SmallPulse}

Whereas the previous calculation was merely a warm--up exercice, which
permitted us to fix the notations and introduce some general formul\ae,
the calculation to follow will provide us with the first non--trivial
example of the phenomenon that we would like to emphasize throughout this
paper: a source which radiates in the vacuum of the strongly coupled
${\mathcal N}=4$ SYM theory produces an energy density which exhibits
{\em no quantum broadening} in the supergravity approximation.

Here, the {\em source} will be the heavy quark which is accelerated under
the action of a weak\footnote{The case of a large, but uniform, angular
acceleration has been studied in Ref.~\cite{Athanasiou:2010pv} and will
be also discusses in Sect.~\ref{Speed} below; other cases will be
considered in a subsequent paper \cite{wip}.} but otherwise arbitrary
external force, and thus can radiate quanta of ${\mathcal N}=4$ SYM. From
the perspective of the dual, gravity, calculation, this means that the
endpoint of the string at the boundary is forced to follow some arbitrary
motion, whose amplitude is weak and slowly varying in time (see below for
the precise conditions). Via the (linearized) equations of motion, this
perturbation of the endpoint propagates as a small perturbation of the
shape of the string inside the bulk.

Furthermore, the {\em lack of quantum broadening} means that the energy
density {\em radiated by the quark} --- {\em i.e.}, the backreaction
produced by the string perturbation on the boundary --- exhibits the
same, localized, distribution in space and time as the corresponding
solution of the classical Maxwell equations: the radiated energy appears
to propagate at the 4D speed of light, so like classical fields or free
massless quanta. As we shall see, the absence of broadening is a
consequence of the fact that the string backreaction responsible for
radiation is restricted to the {\em string endpoint} at $z=0$ alone.

For simplicity, we shall assume that the string is at rest in the absence
of the external perturbation and that its deformation from a vertical
line is restricted to just one spatial direction: the $x\equiv x^1$ axis.
(If the latter condition is satisfied by the string endpoint, then it is
automatically satisfied by the string perturbation anywhere in the bulk.)
We shall present explicit results for arbitrary small perturbations, but
the physically most transparent example is that of an external force
which is localized in time --- a `kick' acting on the heavy quark. The
associated string perturbation is a pulse which propagates down the
string. The corresponding backreaction has been already computed in
Ref.~\cite{Callan:1999ki}, but only insofar as the `dilaton' field (the
expectation value of the operator ${\rm Tr}\,F_{\mu\nu}^2$) is concerned.
As we shall later explain, that particular case is rather peculiar and in
any case very different from the backreaction for the energy density.

Consider therefore a string which in the absence of any external
perturbation is `sitting' at $\br=0$. (The corresponding formul\ae{} are
obtained by letting $\upsilon\to 0$ in all the formul\ae{} in
Sect.~\ref{Constant}.) Under the action of an external force, the string
endpoint acquires a time--dependent deviation $x_q(t)$ along the $x$
axis, which is `small' in a sense to be shortly specified. Then the
general perturbation of the string can be parametrized as $x_s(t,z)$ with
boundary condition $x_s(t,z=0) = x_q(t)$. The string embedding functions
are $X^M = (t,x_s,0,0,z)$ and the Nambu--Goto action takes the form
 \beq
 S = - T_0 \int \dif t\, \dif z\, \sqrt{-g} =
 -T_0 \int \dif t\, \dif z |G_{00}| \sqrt{1-\dot{x}_s^2+x_s^{\prime 2}}
 \eeq
with $G_{00} = -L^2/z^2$. Varying $x_s \to x_s + \delta x(t,z)$ and
requiring the action to be stationary we obtain the classical string
equation of motion (EOM)
 \beq\label{EOM}
 \frac{\del }{\del t}\,\frac{\dot{x}_s}{\sqrt{1-\dot{x}_s^2+x_s^{\prime 2}}}
 - \frac{1}{|G_{00}|}\,\frac{\del}{\del z} \, \frac{|G_{00}| x'_s}{\sqrt{1-\dot{x}_s^2+x_s^{\prime 2}}} = 0.
 \eeq
Assuming that $|\dot{x}_s|, |x'_s| \ll 1$ the EOM linearizes in $x_s$ and
becomes
 \beq\label{EOM linear}
 \left(\del_t^2 -\del_z^2 + \frac{2}{z}\,\del_z \right) x_s = 0.
 \eeq
The general retarded solution to \eqnum{EOM linear} with the given
boundary condition is
 \beq\label{xs linear}
 x_s(t,z) = x_q(t-z) + z \dot{x}_q(t-z).
 \eeq
Notice that this is the linear approximation to the general solution of
the EOM \cite{Mikhailov:2003er}
 \beq\label{string general}
 t = t_q + \gamma_q  z,
 \qquad
 x_s = x_q + \gamma_q \dot{x}_q z,
 \eeq
where $x_q, \dot{x}_q$ and $\gamma_q$ are evaluated at $t_q$.
Interestingly, the linear approximation is tantamount to letting
$\gamma_q \to 1$, that is, it is tantamount to the non--relativistic
approximation for the motion of the quark on the boundary. In general, it
is helpful to think that the boundary motion can be parametrized as
 \beq\label{xq generic}
 x_q = x_0 f(t/\tau, \dots),
 \eeq
where $\tau$ is the smallest time scale in the problem and the dots stand
for the dependence on the remaining time scales. Then one can see that
the linear approximation holds under the conditions that $x_0 \ll \tau$
and $z \ll \tau^2/x_0$. Thus, even for boundary perturbations whose
amplitude $x_0$ is arbitrarily small, the linear approximation can be
trusted only up to some maximal penetration $z$ inside the bulk, beyond
which the string perturbation (which is amplified by the scale factor
$z$) becomes relatively large.

To be more specific, let us consider a perturbation which is localized in
time :
  \beq\label{GaussString}
 x_q = x_0 \exp \left(-\frac{t^2}{2\sigma^2} \right)\,.
 \eeq
In the linear approximation, this leads to the following string profile
 \beq\label{Pulse}
 x_s = x_0 \left[1 + \frac{z(z-t)}{\sigma^2}\right]
 \exp\left[-\frac{(z-t)^2}{2 \sigma^2}\right],
 \eeq
which describes a pulse on the string with a maximum at $z=t$ and a width
of order $\sigma$. Hence, the position of the peak propagates down the
string at the 5D speed of light: $v_z=\dif z/\dif t = 1$. The linear
approximation is valid as long as $x_0 \ll \sigma$ and $z\ll
\sigma^2/x_0$; these conditions allow for values of $z$ which are much
larger than $\sigma$.

Returning to the general perturbation \eqref{xs linear}, let us compute
the {\em radiated} energy density, as produced via backreaction on the
boundary. To that aim, we shall rely on the general formul\ae{}
introduced in Sect.~\ref{Constant} and in particular on \eqnum{tMN} for
the string stress tensor, where now $\delta^{(3)}(\br - \br_s) =
\delta^{(2)}(\bxT) \delta(x - x_s)$. In what follows, we shall simplify
these formul\ae{} in order to \texttt{(i)} be consistent with the linear
approximation for the string perturbation, and \texttt{(ii)} retain only
those contributions which correspond to radiation.

The second constraint above turns out to be quite subtle: the string
worldsheet is simultaneously describing various forms of energy --- the
Coulomb energy of the heavy quark, the radiated energy, and interference
terms between the two --- and in general it does not seem possible to
unambiguously isolate these various contributions already at the level of
the string stress tensor. Rather, the general strategy in that sense is
to first compute the total energy density on the boundary (via
backreaction from the string in the bulk) and then identify the radiation
as the component of the energy density which shows the slowest fall--off,
namely like $1/r^2$, when $r\to\infty$. In our present analysis, we shall
partly follow this strategy --- by chasing contributions with a slow
fall--off at large $r$ ---, but we shall combine it with a physical
analysis of the string stress tensor $t_{MN}$, which will allow us to
identify and discard the contributions to the Coulomb energy.

For more clarity, we relegate most of the technical details to Appendix
\ref{appPulse} and focus here on the main results and on the points of
physics. A first conclusion of the analysis in Appendix \ref{appPulse} is
that, to the accuracy of interest, one can ignore the small deviation
$x_s$ away from $x=0$ within the $\delta$--function expressing the
support of the string; that is, one can replace $\delta^{(3)}(\br -
\br_s)\,\to\,\delta^{(3)}(\br)$. Indeed, the contributions obtained by
expanding $\delta(x - x_s)$ in powers of $x_s$ represent, at most,
interference effects between Coulomb energy and radiation and thus decay
as $1/r^3$ or faster at large distances\footnote{Note that such an
interference term represents the leading order correction to the
`dilaton' field due to the static string \cite{Callan:1999ki}. So, in
that case, it was important to keep trace of $x_s$ inside $\delta(x -
x_s)$.}. We therefore write
 \beq\label{tmnPulse}
 t_{MN} = \tilde{t}_{MN} \delta^{(3)}(\br)\,.
 \eeq
The various components $\tilde{t}_{MN}$ are listed in \eqnum{tmn string}
in Appendix \ref{appPulse}. Here, we focus on the energy density
$\tilde{t}_{MN}$, which reads:
 \beq\label{t00}
 \tilde{t}_{00} \,=\,
 \frac{\sqrt{\lambda}}{2 \pi}\, \frac{z}{L^3}\,
 \frac{1 + x_s^{\prime 2}}{\sqrt{1-\dot{x}_s^2+x_s^{\prime 2}}}\,,\qquad
 E_s\,=\,\int_{z_m}^\infty \dif z\,
 \frac{L^3}{z^3}\ \tilde{t}_{00}(t,z)\,. \eeq
Given $\tilde{t}_{00}$, the total energy $E_s$ stored in the string is
computed as shown above. As anticipated, $\tilde{t}_{00}$ also encodes
the Coulomb energy, which would be non--vanishing even in the absence of
acceleration. For instance, for a heavy quark with constant velocity
$v^1=\upsilon$ (the problem discussed in Sect.~\ref{Constant}), we have
$\dot{x}_s=\upsilon$ and $x'_s=0$, and the above equations yield $E_s=m_q
\gamma$, as expected for a relativistic particle. (Recall that $z_m=
{\sqrt{\lambda}}/(2\pi m_q)$ with $m_q$ the mass of the quark.) For a
generic motion of the heavy quark, \eqnum{t00} simultaneously describes
Coulomb energy (associated with the {\em instantaneous} value of
$\dot{x}_s$) and radiation (associated with the {\em variation} of
$\dot{x}_s$, {\em i.e.} with the quark acceleration, and --- related to
this --- to the {\em deformation} of the string worldsheet, as measured
by $x'_s$). As we now demonstrate, these two types of energy can be
disentangled from each other for the small perturbation problem at hand.

For consistency with the linear approximation in \eqref{xs linear}, we
need to expand \eqnum{t00} to quadratic order in the perturbation $x_s$.
This yields
 \beq\label{t002nd}
 \tilde{t}_{00} \,\simeq\,
 \frac{\sqrt{\lambda}}{2 \pi}\, \frac{z}{L^3}\,
 \left(1+ \frac{1}{2}\, x_s^{\prime 2} + \frac{1}{2}\,\dot{x}_s^2 \right)
 .\eeq
At this level, it is convenient to use \eqnum{xs linear} in order to
relate $\dot{x}_s$ and $x_s'$ to the boundary motion :
 \beq\label{xs der}
 \dot{x}_s = \dot{x}_q + z \ddot{x}_q,
 \qquad x_s' = -z \ddot{x}_q,
 \eeq
where in the r.h.s. $x_q$ is evaluated at $t-z$. By inserting these
expressions in \eqnum{t002nd} and using the fact that  $\dif/\dif t =
-\dif /\dif z$ when acting on $x_q$, we deduce
 \beq\label{t00xq}
 \tilde{t}_{00} &\,\simeq\,&
 \frac{\sqrt{\lambda}}{2 \pi}\, \frac{z}{L^3}\,
 \left[1 +\frac{1}{2}\,
 \left(\dot{x}_q^2 + 2z\dot{x}_q\ddot{x}_q\right)
 +\,z^2\ddot{x}_q^2\right]\nn
 &\,=\,&
 \frac{\sqrt{\lambda}}{2 \pi}\, \frac{z^3}{L^3}\,
 \left[\frac{1}{z^2} - \frac{1}{2}\,\frac{\dif}{\dif z}
 \left(\frac{\dot{x}_q^2}{z}\right)+\,\ddot{x}_q^2\right].
 \eeq
We shall now argue that the first two terms within the last square
bracket, that is  $1/z^2$ and the term which is a total derivative,
represent the Coulomb energy of the non--relativistic quark, whereas the
last term, proportional to the quark acceleration squared, represents the
radiation. To that aim, we compute the total string energy according to
\eqnum{t00} :
\beq E_s&\,\simeq\,&\frac{\sqrt{\lambda}}{2 \pi}
 \int_{z_m}^\infty\dif z\,\left[\frac{1}{z^2} - \frac{1}{2}\,
 \frac{\dif}{\dif z}
 \left(\frac{\dot{x}_q^2}{z}\right)+\,\ddot{x}_q^2\right]
 \nn &\,=\,&
 \frac{\sqrt{\lambda}}{2 \pi z_m}\left(1+\frac{\upsilon^2}{2}
 \right)\,+\frac{\sqrt{\lambda}}{2 \pi}
 \int_{z_m}^\infty\dif z\,\ddot{x}_q^2(t-z)\nn
 &\,=\,& m_q +\,\frac{m_q \upsilon^2}{2} +\,\frac{\sqrt{\lambda}}{2 \pi}
 \int_{-\infty}^{t} \dif \tp\, \ddot{x}_q^2(\tp),
 \eeq
where $\upsilon=\dot{x}_q(t)$. As anticipated, the first two terms in the
last expression are recognized as the quark energy in the
non--relativistic approximation, whereas the remaining integral involving
$\ddot{x}_q^2$ is clearly the radiated energy.

Incidentally, the above calculation already tells us what should be the
total energy radiated by the quark: by energy conservation, this is the
same as the radiative piece of the energy stored into the string. Hence,
 \beq\label{etot}
 E_{\rm rad}\, =\,\frac{\sqrt{\lambda}}{2 \pi}
 \int_{-\infty}^{t} \dif \tp\, \ddot{x}_q^2(\tp),\qquad
 P\,\equiv\,\frac{\dif E_{\rm rad}}{\dif t} = \frac{\sqrt{\lambda}}{2
 \pi}\,a^2\,,
 \eeq
where $a=\ddot{x}_q$ is the quark acceleration and $P$ is the radiated
power. These formul\ae{} were to be expected: as shown by Mikhailov
\cite{Mikhailov:2003er}, they hold for an arbitrary motion of the quark
on the boundary. Moreover, they are formally identical with the
corresponding classical result in electrodynamics, up to the replacement
$\sqrt{\lambda}\leftrightarrow e^2/3$. Recovering these results via the
explicit calculation of the string backreaction represents a non--trivial
check on the respective calculation, to which we now return.

Via similar manipulations, one can convince oneself that the radiative
contributions to all the components of $\tilde{t}_{MN}$ are proportional
to $a^2$ (to the accuracy of interest). The respective results are quite
simple: the only non--trivial components are
 \beq\label{tradPulse}
 \tilde{t}_{00}^{\rm rad} = \tilde{t}_{55}^{\rm rad}
 = -\tilde{t}_{05}^{\rm rad} =
 \frac{\sqrt{\lambda}}{2 \pi}\,\frac{z^3}{L^3}\,\ddot{x}_q^2
 \eeq
with $\ddot{x}_q^2$ evaluated at $t-z$. Using these results, one can
evaluate the radiated energy density according to Eqs.~\eqref{EA} and
\eqref{EB}. It turns out that $\EB=0$ whereas 
 \beq\label{EAPulse}
 \EA = \frac{\sqrt{\lambda}}{\pi^2}\,
 \del^2_{r^2}
 \int \dz\, \dif \tp\,\delta(\mcal{W}) \,z (z+ t-\tp)\, \ddot{x}_q^2
 (\tp-z)\,.
 \eeq
The linear approximation is {\em a priori} restricted to relatively small
values of $z$, but this poses no problem for the above integral over $z$,
since this is actually saturated by its lower limit at $z_m\approx 0$
(see below). Also, $\mcal{W}$ has to be evaluated at $x_s=0$, cf.
\eqnum{tmnPulse}, and therefore (for the retarded solution)
 \beq
 \delta(\mcal{W}) =
 \frac{\delta(\tp - t + \sqrt{z^2+r^2})}{2 \sqrt{z^2 + r^2}}.
 \eeq
By using this $\delta$--function to perform the time integration, we
obtain
 \beq\label{EintPulse}
 \mcal{E} = \frac{\sqrt{\lambda}}{2 \pi^2}\,
 \del^2_{r^2}
 \int \dz \,
 \frac{z (z+ \sqrt{z^2 + r^2})}{\sqrt{z^2 + r^2}}\,
 \ddot{x}_q^2.
 \eeq
with $\ddot{x}_q^2$ evaluated at $\tp-z = t -z-\sqrt{z^2+r^2}$. At this
point, it is convenient to change the integration variable from $z$ to
$\xi$, via
\beq\label{xi}
 \xi \equiv z + \sqrt{z^2+r^2},
 \eeq
This implies \beq
 z = \frac{\xi^2 - r^2}{2 \xi}, \qquad \dif z = \frac{\xi^2 + r^2}{2 \xi^2}\, \dif \xi, \qquad \sqrt{z^2+r^2}=\frac{\xi^2 + r^2}{2 \xi},
 \eeq
and the energy density becomes (at the lower limit, we neglect $z_m$ next
to $r$)
 \beq\label{EdPulse}
 \mcal{E}\, =\, \EA\,=
 \frac{\sqrt{\lambda}}{4 \pi^2}\, \del^2_{r^2}
 \int_r^{\infty}
 \frac{\dif \xi}{\xi}\,
 (\xi^2 - r^2) \ddot{x}_q^2\,,
 \eeq
where now the argument of $x_q$ is $t - \xi$ and every dot corresponds to
a derivative w.r.t. $t$ or equivalently the argument.

At this point, an essential simplification occurs: since the integrand in
\eqnum{EdPulse} is linear in $r^2$, it is quite clear that, after we take
the external derivatives, the only surviving contribution is a {\em
boundary term}, coming from the lower boundary at $\xi=r$, or $z=0$. A
similar property was noticed in Ref.~\cite{Athanasiou:2010pv} in relation
with the rotating string. On the other hand, the situation is very
different for the calculation of the average `dilaton' density
$\langle{\rm Tr}\,F_{\mu\nu}^2\rangle$ on the boundary, where a final
integration over all values of $\xi$ is left to be done
\cite{Callan:1999ki}. This difference has dramatic consequences for the
space--time pattern of the response on the boundary, which we shall
shortly discuss.

Returning to \eqnum{EdPulse}, it is important to stress that the
simplification alluded to above has occurred because of compensations
between the various terms in \eqnum{EA}, which in turn were made possible
by the particularly simple structure of the radiation piece of the bulk
stress tensor, \eqnum{tradPulse}. (For instance, if one takes just the
$\tilde{t}_{00}$ piece in the integrand of \eqnum{EA} and then uses the
simple expression for $\tilde{t}_{00}$ shown in \eqnum{tradPulse}, one
obtains terms $\propto r^4$ which would yield non--trivial contributions
from the bulk.) We conclude that both the particular tensor structure in
\eqnum{tradPulse}, and the specific functional form of the individual
components there, are important for our final conclusion that only
boundary terms survive. These boundary terms are easily evaluated as
\beq\label{ed xq}
 \mcal{E}(t,r) = \frac{\sqrt{\lambda}}{8 \pi^2}\, \frac{\ddot{x}_q^2(t-r)}
 {r^2}\,.
 \eeq
The total radiated energy is obtained by multiplying the above expression
with $4\pi r^2$ and integrating over $r$. Clearly, this yields the
expected result, cf. \eqnum{etot}.

From the previous discussion, it should be clear that \eqnum{ed xq}
represents the general solution for an arbitrary one--dimensional motion
of the heavy quark in the non--relativistic limit. In particular, for the
Gaussian pulse \eqref{Pulse} one finds
 \beq\label{EPulse}
 \mcal{E} = \frac{\sqrt{\lambda}}{8 \pi^2}\,\frac{x_0^2}{\sigma^8 r^2}\,
 \left[(r - t)^2 - \sigma^2 \right]^2 \exp\left[-\frac{(r -t )^2}{\sigma^2}
 \right],\qquad E_{\rm rad}
 = \frac{3 \sqrt{\lambda}}{8 \sqrt{\pi}}\,\frac{x_0^2}{\sigma^3}.
 \eeq

The most important feature of \eqnum{ed xq} for us here is the fact that
the space--time distribution of the radiated energy is controlled by the
function $x_q(t-r)$, {\em i.e.} by the initial perturbation of the heavy
quark translated from $r=0$ (the quark position) to $r=t$. A similar
pattern would be obtained by solving the classical Maxwell equations with
the heavy quark as a source. It looks like, at strong coupling, the
radiation is simply propagating from $r=0$ up to $r=t$ at the speed of
light, in the same way as classical radiation (or free massless quanta)
would do\footnote{Although similar to a classical solution insofar as the
space--time localization is concerned, the SUGRA result \eqref{ed xq}
differs from the corresponding classical result in some other important
features, like the fact that it is isotropic and the proportionality with
$\sqrt{\lambda}$.}. In particular, for the localized perturbation
\eqref{GaussString}, the radiated energy density, \eqnum{EPulse},
propagates as a spherical shell centered at $r=t$ and with a width equal
to the width $\sigma$ of the initial perturbation. There is no sign of
{\em quantum broadening}, that is, no time--like components in the
radiation left behind at $r \ll t$, in contrast to expectations based on
the physical picture of parton branching at strong coupling (see the
discussion at the beginning of Sect.~\ref{sect-falling}).

In what follows, we shall argue that this lack of broadening of the
radiative energy density on the boundary is a consequence of the fact
that the whole backreaction comes from the string endpoint at $z=0$, as
noticed after \eqnum{EdPulse}. To that aim, we shall show that the
backreaction from string points at $z>0$ --- if non--zero ! --- would
introduce a spread $t-r\sim z$ in the energy distribution, which with
increasing $z$ could become arbitrarily large. For more clarity we shall
focus on a localized initial perturbation, like \eqnum{GaussString}. Then
the perturbation propagates down the string as a pulse centered at
$z=\tp$ (cf. \eqnum{Pulse}). This property together with the causality
condition $\mcal{W}=0$ implies the following relation between the
position $z$ of the `source in the bulk' (here, a point on the pulse) and
the measurement point $(t,r)$ on the boundary: $t\simeq z +
\sqrt{z^2+r^2}$ or
 \beq\label{UVIRPulse}
  z\,\simeq\,\frac{t^2-r^2}{2t}\,\simeq\frac{t-r}{2}\,,\eeq
where the second approximation holds when both $t$ and $r$ are large with
$t,\,r\gg t-r$. This condition is easy to understand: it takes a time
$\tp\simeq z$ for the perturbation to propagate along the string from
$z=0$ up to $z$ and then a time $\sqrt{z^2+r^2}$ for the gravitational
wave expressing the backreaction to propagate through AdS$_5$ from the
point $z$ on the string (which has $r=0$) up to the point $r$ on the
boundary. Note that both propagations alluded to above proceed at the 5D
speed of light.

\eqnum{UVIRPulse} is the expected form of the UV/IR correspondence ---
or, more precisely, its expression \texttt{(ii)} according to the
discussion in the Introduction. It predicts a spreading $t-r$ in the
energy distribution which grows proportionally with the radial location
$z$ of the source in the bulk. Such a spreading is seen indeed in the
calculation of the `dilaton' density on the boundary
\cite{Callan:1999ki}, but it is absent from the corresponding
calculations of the radiation, as presented here and in
Ref.~\cite{Athanasiou:2010pv}. Clearly, this absence of broadening is
still consistent with \eqnum{UVIRPulse} because in the case of radiation
the whole backreaction comes from the endpoint of the convolution at
$z=z_m$. Yet, this is a rather surprising feature, that we would like to
understand better via the remaining analysis in this paper.


\subsection{Velocity of a bit of energy}
\label{Speed}

In this subsection, we start developing an argument, to be completed at
later stages of this analysis, which sheds more light on the absence of
broadening in the SUGRA calculations of radiation. Namely, we shall argue
that the lack of backreaction from bulk sources which lie far away from
the boundary is a consequence of the fact that these sources propagate at
the speed of light in AdS$_5$. We should stress from the very beginning
that this property is not just a consequence of the kinematics, and hence
of causality. Indeed, we have just seen, towards the end of the previous
subsection, that causality alone does permit backreaction (in terms of
radiated energy on the boundary) from bulk points at $z>0$ and that the
respective contributions would show spreading, cf. \eqnum{UVIRPulse}. The
fact that such contributions are nevertheless absent in the final result
is therefore a property of the SUGRA dynamics, that we shall now
attribute to the propagation of bulk sources at the 5D speed of light.

The argument will be developed in several steps: First, in the present
subsection, we shall verify that the flow of energy in AdS$_5$ proceeds
indeed at the speed of light (except near $z=0$) for the two examples
that we have discussed so far: the small string perturbation and the
rotating string of Ref.~\cite{Athanasiou:2010pv}. Then, in
Sect.~\ref{Falling} we shall show that this property --- the propagation
of bulk sources at the speed of light --- implies the particular
structure for the bulk stress tensor shown in \eqnum{tradPulse}, which in
turn is responsible for the lack of backreaction (as discussed after
\eqnum{EdPulse}). Finally in Sect.~\ref{sect-LC} we shall argue that the
fact bulk sources propagating at the speed of light cannot radiate may be
interpreted as a form of Lorentz time dilation, but in a `twisted' system
of coordinates which mixes the 5th dimension with one of the spatial
coordinates $x^i$ on the boundary.

In the context of Sect.~\ref{SmallPulse}, the fact that the perturbation
propagates down the string at the speed of light seems already well
established: we have noticed e.g. that the peak of the pulse in
\eqnum{Pulse} travels according to $z=t$. However, other examples like
the rotating string \cite{Athanasiou:2010pv} or the trailing string (at
finite temperature) \cite{Herzog:2006gh,Gubser:2006bz} show that it is
not always possible to associate the flow of energy along a string with a
propagating string deformation. So, we need a more precise definition of
what we mean by the `velocity of the energy flow down the string'. This
is provided by the string stress tensor $t^{MN}$: the component $t^{0M}$
(with $M\ne 0$) describes the energy flow in the $M$ direction, while
$t^{00}$ is the energy density; hence the ratio
$\upsilon_M=t^{0M}/t^{00}$ defines the respective component of the
velocity\footnote{Note that $\upsilon_M$ is not a 5D vector, so the
position of the index $M$ is irrelevant.}. As explained in
Sect.~\ref{SmallPulse}, care must be taken to include into $t^{0M}$ and
$t^{00}$ only the respective radiative contributions, since the Coulomb
energy is not flowing. We shall therefore use
 \beq\label{vm0}
 \upsilon_M\,=\,\frac{t^{0M}_{\rm rad}}{t^{00}_{\rm rad}}\,.\eeq
For instance, this definition together with \eqnum{tradPulse} immediately
implies $\upsilon_z=1$ (and of course $\upsilon_i=0$ for $i=1,2,3$), at
any $z>0$. So, for that problem, we have $\upsilon^2=\upsilon_z^2=1$, as
anticipated.

An alternative definition, which is equivalent to \eqnum{vm0} but often
easier to use in practice, involves the worldsheet stress tensor
$\pi_M^a$ with $a=\tau,\,\sigma$. Within the string worldsheet, there is
only one direction for the energy flow, namely $\sigma=z$. (As before,
$\tau=t$ and $\sigma=z$.) Then it is natural to define the radial
velocity of a bit of energy as
 \beq\label{vs0}
 \upsilon_z \,\equiv\, \frac{\dif E_s/ \dif t}
 {\dif E_s/ \dif z}\bigg|_{\rm rad}\,,
 \eeq
where $\dif E_s/ \dif z$ and $\dif E_s/ \dif t$ denote the energy density
and the energy flow down the string, respectively, and the subscript
``rad'' indicates that we have to consider only radiative contributions
to these quantities. One generally has (with $T_0 = \sqrt{\lambda}/2 \pi
L^2$)
 \beq\label{es den}
 \frac{\dif E_s}{\dif z} &\,=\,& - \pi_0^{\tau} =
 \frac{T_0}{\sqrt{-g}}\,[(\dot{X} \cdot X')X'_0 - (X' \cdot
 X')\dot{X}_0]\,,
 \nn
 \frac{\dif E_s}{\dif t} & \,=\,& - \pi_0^{\sigma} =
 \frac{T_0}{\sqrt{-g}}\,[(\dot{X} \cdot X')\dot{X}_0 -
 (\dot{X} \cdot \dot{X})X'_0]\,.
 \eeq
One way to understand \eqnum{vs0} is to follow a curve of constant energy
$E_s(t,z)$. The condition $\dif E_s=0$ under small variations $\dif t$
and $\dif z$ implies indeed
 \beq\label{vs1}
 \dif E_s\,=\,\frac{\del E_s}{\del t}\,\dif t
 +\frac{\del E_s}{\del z}\,\dif z\,=\,0\ \Longrightarrow\
 \upsilon_z\,\equiv\,\frac{\dif z}{\dif t}\,=\,\frac{-{\del E_s}/{\del t}}
 {{\del E_s}/{\del z}}\,.\eeq
The difference in sign w.r.t. \eqnum{vs0} comes from the fact that, in
\eqnum{es den}, we have defined the energy flux to be positive when the
energy is decreasing at a given $z$.

\comment{\beq\label{delinear}
 \frac{\dif E_s}{\dif z}  =
 \frac{\sqrt{\lambda}}{2 \pi z^2}\,
 \frac{1 + x_s^{\prime 2}}{\sqrt{1-\dot{x}_s^2+x_s^{\prime 2}}},\qquad
 \frac{\dif E_s}{\dif t} = -\frac{\sqrt{\lambda}}{2 \pi z^2}\,
 \frac{\dot{x}_s x'_s}{\sqrt{1-\dot{x}_s^2+x_s^{\prime 2}}}.
 \eeq
 }

It is straightforward to apply Eqs.~\eqref{vs0}--\eqref{es den} to the
problem of a small string perturbation in Sect.~\ref{SmallPulse} and thus
rederive the expected result $\upsilon_z=1$. (The separation of the
radiated energy within the worldsheet tensor proceeds in exactly the same
way as for the respective 5D stress tensor, cf. the discussion leading to
\eqnum{tradPulse}.) Here we shall rather use them for the problem of the
rotating string, {\em i.e.} a string attached to a heavy quark which
rotates on the boundary at uniform angular velocity $\omega_0$. The
corresponding backreaction has been computed in
Ref.~\cite{Athanasiou:2010pv} and found to arise from the string endpoint
at $z=0$ alone. In agreement with that finding, we shall now show that
the velocity of a bit of energy flowing down the string is equal to one
at any $z>0$. In spherical coordinates, the profile of the rotating
string reads \cite{Athanasiou:2010pv}
 \beq\label{xsrot}
 r_s = \textstyle{\sqrt{R_0^2 + \gamma_0^2 \upsilon_0^2 z^2}},
 \qquad
 \theta_s = \displaystyle{\frac{\pi}{2}},
 \qquad
 \phi_s = \omega_0 (t - \gamma_0 z) + \arctan(\gamma_0 \omega_0 z),
 \eeq
where $\omega_0 = \upsilon_0/R_0$, $\gamma_0^2 = 1/(1-\upsilon_0^2)$ and
with $\upsilon_0$ the magnitude of the boundary velocity. Inserting these
formul\ae{} into \eqnum{es den}, it is straightforward to deduce that
 \beq\label{ed rotating}
 \frac{\dif E_s}{\dif z} &=& \frac{\sqrt{\lambda}}{2 \pi}\,\frac{\gamma_0}{z^2} +
 \frac{\sqrt{\lambda}}{2 \pi}\, \gamma_0 a_0^2,
 \\ \label{ef rotating}
 \frac{\dif E_s}{\dif t} &=& \frac{\sqrt{\lambda}}{2 \pi}\,a_0^2.
 \eeq
Here $a_0 = \gamma_0^2 \upsilon_0^2/R_0$ is the proper acceleration of
the boundary motion, which also determines the position of the induced
worldsheet horizon: $z_{\rm h} = 1/a_0$. Considering the energy density
in \eqnum{ed rotating}, it seems natural to associate the first term with
the Coulomb energy density $\dif E_{\rm Coul}/\dif z$ of the rotating
quark, and the second one with the energy density due to radiation, $\dif
E_{\rm rad}/\dz$. This identification is supported by the type of
arguments leading to \eqnum{tradPulse}: after integrating \eqnum{ed
rotating} over $z$ to compute the total energy stored in the string, one
finds that the first term there yields a contribution $m_q\gamma_0$,
which is recognized as the energy of a relativistic quark with uniform
velocity $v_0$. As for the energy flux in \eqnum{ef rotating}, this
coincides with the expected result for radiated power, so it is clear
that this includes radiation alone. Hence, the radial velocity of a bit
of energy is given by
 \beq\label{vz}
 \upsilon_z&=&\frac{\dif E_s/ \dif t}{\dif E_{\rm rad}/ \dif z}
 = \frac{1}{\gamma_0}\,.\eeq
This is strongly suppressed in the ultrarelativistic limit $\gamma_0\gg
1$, which can be interpreted as a consequence of the Lorentz collimation
of the radiated energy around the direction of emission (here, in the
five--dimensional sense). But a small progression in $z$ along the
profile of the string implies a relatively large displacement in the
spatial direction $r$, since the respective coordinate $r_s$ of the
string rises rapidly with $z$, cf. \eqnum{xsrot}. Specifically, the
energy bit has a non--zero velocity $v_r$, computed as
 \beq\label{vr}
 \upsilon_{r}\equiv \frac{\dif r_s}{\dif t}
 =\dot{r}_s + r'_s \upsilon_z
 = \frac{\gamma_0 \upsilon_0^2 z}{\sqrt{R_0^2+\gamma_0^2 \upsilon_0^2 z^2}}
 \,,
 \eeq
with of course $\dot{r}_s=0$ for the rotating string. Finally, the string
as a whole is rotating, and therefore so does the energy bit flowing
along it. This implies the following $\upsilon_{\phi}$ component for the
velocity of the energy flow
 \beq\label{vphi}
 \upsilon_{\phi}\equiv r_s\frac{\dif \phi_s}{\dif t}
 =r_s \dot{\phi}_s + r_s \phi'_s \upsilon_z
 =\frac{\upsilon_0 R_0}{\sqrt{R_0^2+\gamma_0^2 \upsilon_0^2 z^2}}\,.
 \eeq
(The other angular velocity vanishes, $\upsilon_{\theta} = 0$, since
$\theta$ is fixed for the string.) On the boundary ($z=0$), one has
$\upsilon_{\phi} = \upsilon_0$ with $\upsilon_{r}$ vanishing, while for
large $z$, on the contrary, $\upsilon_{\phi}$ vanishes with $\upsilon_{r}
\to \upsilon_0$. For the magnitude squared we obtain
 \beq
 \upsilon^2 \,\equiv\,\upsilon_{z}^2+\upsilon_{r}^2+\upsilon_{\phi}^2\,=\,
 1,
 \eeq
for any $z$ and for any boundary velocity $\upsilon_0$. As a check,
notice that we would obtain the same results by computing the velocity of
the energy flow directly from the 5D string stress tensor, according to
\eqnum{vm0}. The components of interest are the energy and momentum
density given in spherical coordinates by \cite{Athanasiou:2010pv}
 \beq\label{t0m}
 t^{0M}_{\rm rad} = \frac{\sqrt{\lambda} \gamma_0 z^5}{2\pi L^7}\, \delta^{(3)}(\br-\br_s)
 (r_s^{\prime 2} + r_s^2 \phi_s^{\prime 2}\,,\,
 -\omega_0 r_s^2 r_s' \phi'_s\,,\,
 0\,,\,
 \omega_0 r_s^{\prime 2}\,,\,
 -\omega_0 r_s^2 \phi'_s),
 \eeq
with the elements corresponding to ($t,r,\theta,\phi,z$) respectively and
where we have again subtracted from $t^{00}$ and $t^{0\phi}$ the terms
which correspond to the static energy of the string. (We shall elaborate
more on this subtraction in the subsequent publication \cite{wip}.) Then
the velocity of the energy flow is
 \beq\label{vm}
 \upsilon_{M} = \frac{1}{t^{00}_{\rm rad}}\,\left(t^{0r}_{\rm rad},t^{0\theta}_{\rm rad},r_s t^{0\phi}_{\rm rad},t^{0z}_{\rm rad}\right),
 \eeq
with $M=r,\theta,\phi,z$. Plugging \eqnum{t0m} into \eqnum{vm} we arrive
at the previously found expressions in Eqs.~\eqref{vz}, \eqref{vr} and
\eqref{vphi} for the components of the velocity.

Let us also comment here on some other intriguing result in
Ref.~\cite{Athanasiou:2010pv}, namely the fact that the {\em whole}
contribution to the energy density on the boundary, and not just its
radiative part, appears to be generated by backreaction from the endpoint
of the string at $z=0$. Whereas for radiation this feature can be
ascribed to the propagation of the bits of energy down the string at the
speed of light, there is certainly no such a correlation for the Coulomb
piece of the energy, which is static. Rather, what we would like to argue
here is that, for the case of the Coulomb energy, the respective finding
is merely a consequence of a particular choice made in
Ref.~\cite{Athanasiou:2010pv} for the integration variables in the
convolutions expressing the backreaction. Namely, the Coulomb energy
appears as a {\em bulk} contribution ({\em i.e.}, a result of the
backreaction from string points at $z>0$) when using the standard
variables $\tp$ and $z$ --- in terms of which we have a transparent
physical interpretation and the UV/IR correspondance holds as expected
---, but it can be formally transferred into a {\em boundary} contribution
(a backreaction from $z=0$ alone) via an appropriate change of variables
mixing $\tp$ and $z$. To illustrate this point, we shall perform in
Appendix \ref{appheavy} the corresponding change of variables on the
example of a string (heavy quark) with constant velocity and show that,
with the new variables, the complete result appears to be generated at
the endpoint of the integration at $z=0$. Yet, from the discussion in
Sect.~\ref{Constant}, we already know that string points at any value of
$z$ do contribute to the Coulomb energy, provided one uses the `physical'
variables $\tp$ and $z$. This example shows that care must be taken when
trying to understand the UV/IR correspondence in various sets of
coordinates.

At this point, one may wonder whether a similar property could also hold
for the {\em radiated} energy --- namely, whether the fact that the whole
contribution to radiation is coming from $z=0$ is not just an artifact of
our peculiar choice of coordinates. We do not believe this to be so,
since in this case the correlation with $z=0$ has a clear signature,
which is independent of the choice of integration variables in the
calculation of backreaction: the lack of broadening of the radiated
energy density in physical space. As we saw on the example of the small
string perturbation, the width $t-r$ of the energy distribution on the
boundary, \eqnum{EPulse}, is fixed by the width $\sigma$ of the initial
perturbation. Via the UV/IR correspondence, this allows for contributions
from string points at $z\lesssim 1/\sigma$, but not from much larger
values of $z$. So, whatever choice we make for the integration variables,
the backreaction must come from points $z$ {\em near} the boundary,
although it is likely that our previous respective choices were indeed
special, in the sense that, with these choices, the whole contribution
was generated from $z=0$ {\em alone}.

Let us conclude this section with a comment on the generality of our
conclusions: the correlation between the bulk propagation at the speed of
light and the lack of energy backreaction on the boundary implies that
the absence of broadening for radiation should be a generic feature of
the supergravity calculations. Indeed, whatever is the physical source of
radiation in the dual gauge theory, the corresponding energy flow in
AdS$_5$ will propagate at the speed of light, except possibly for a
transition region at small $z$, where the radiation is still connected to
its source, and whose extent is controlled by the width of the external
perturbation. Hence, the maximal broadening on the boundary will be fixed
by that width as well. In the next section we shall give other examples
of that type, where the bulk excitation is a supergravity field rather
than a Nambu--Goto string.

 \comment{
It is also interesting to consider the total energy radiated from the
perspective of the gravity problem. We shall do this in two ways which
give the same result. Of course the total energy radiated is infinite,
since we have a stationary situation, so we shall put a cutoff $z_{\rm
m}$. First let us integrate the total energy density from the horizon
$z_{\rm h} = 1/a_0$ to the cutoff to get (a cancelation of the two terms
in \eqnum{ed rotating} at the lower limit takes place )
 \beq
 \int_{z_{\rm h}}^{z_{\rm m}} \dif z\,
 \frac{\dif E}{\dif z} = \frac{\sqrt{\lambda}}{2 \pi}\,\gamma_0 \left(a_0^2 z_{\rm m} - \frac{1}{z_{\rm m}} \right).
 \eeq
But this should be equal to the integral from $0$ to $z_{\rm m}$ of the
string energy density due to acceleration which is equal to
 \beq
 \int_{0}^{z_{\rm m}} \dif z\,
 \frac{\dif E_{\rm a}}{\dif z} = \frac{\sqrt{\lambda}}{2 \pi}\,\gamma_0 a_0^2 z_{\rm m},
 \eeq
and in fact it is so long as $z_{\rm m} \gg z_{\rm h}$.
 }

\section{A time--like wave--packet}
\label{sect-falling} 

In this section, we shall consider a different type of radiation, that
emitted by a time--like (or `massive') wave--packet, which decays into
the massless quanta of ${\mathcal N}=4$ SYM. The prototype of this
phenomenon is the evolution of the time--like photon produced in $e^+e^-$
annihilation. To lowest order in perturbation theory at weak coupling,
the photon decays into a pair of quanta (say, a quark and an antiquark)
which in the center of mass frame propagate with equal but opposite
momenta, at the speed of light. The final state in this approximation is
simply a pair of back--to--back partons (`two jets'). If one includes
higher order corrections (again, at weak coupling), then the pair of
partons initially produced by the decay of the virtual photon are
themselves off--shell (time--like) and can radiate other quanta. Then the
final state consists in three or more partons. The stronger the coupling,
the larger is the probability of branching in the final state, and the
more complex is the structure of the latter in terms of partons. In the
strong coupling limit, we expect this branching to be so efficient that
the energy--momentum distribution in the final state --- consisting in
myriads of partons --- is essentially isotropic (in the center of mass of
the virtual photon, of course). This expectation has been confirmed by
the explicit AdS/CFT calculation of the angular correlations of the
energy density in the final state, by Hofman and Maldacena
\cite{Hofman:2008ar}, which revealed that there are no such correlations
at all: the final state is fully isotropic, in a given event.

The physical picture of parton branching alluded to above has also
another implication: it suggests that the energy--momentum distribution
in the final state at strong coupling must be {\em time--like}. Namely,
one expects the energy density to be non--zero essentially everywhere
inside a three--dimensional sphere with radius $r=t$. (We assume that the
initial photon wave--packet was localized near $r=0$.) Indeed, most of
the quanta produced in the intermediate stages of the parton cascade are
time--like and hence propagate with velocities smaller than one, thus
yielding a tail in the energy distribution at $r < t$. This is to be
contrasted with the energy distribution produced by a classical source
localized at $t=0$ and $r=0$, which is a narrow spherical shell with
$r=t$.

Yet, as we shall shortly see, the AdS/CFT calculation of the energy
distribution produced by the time--like (TL) wave--packet at strong
coupling (in the supergravity approximation) produces a narrow spherical
shell which propagates outwards at the (4D) speed of light, so like a
classical perturbation ! The mathematical origin of this result is the
same as for the string perturbation discussed in the previous section:
the supergravity field describing the AdS$_5$ perturbation induced by the
TL wave--packet propagates into the bulk at the 5D speed of light and
therefore has no backreaction in terms of energy density on the boundary.
The latter is fully generated at early times, when the perturbation is
close to the boundary (within a distance set by the width of the initial
wave--packet) and its velocity is still smaller than one.

Whereas the mathematics of the backreaction calculation is quite similar
to that of the string perturbation in Sect.~\ref{SmallPulse}, the
physical interpretation is perhaps sharper for the case of the TL
wave--packet. This will allow us to better appreciate the inconsistency
of the supergravity result with respect to general expectations from
quantum mechanics.

In Sect.~\ref{SmallPulse} we have seen that the small pulse propagating
down a vertical string behaves very much like a massless classical
particle falling into AdS$_5$ at the speed of light. In Sect.~\ref{TLWP}
below, we shall discover that a similar picture holds also for the TL
wave--packet. We therefore begin our discussion in this section by
computing the backreaction due to a classical, massless, particle which
undergoes free motion in AdS$_5$.

\subsection{The falling massless particle}
\label{Falling}

The motion of a free particle in a curved space--time follows a geodesic.
For a massless particle in AdS$_5$ and with our choice \eqref{metric} for
the metric, the most general such trajectory can be cast into the form
 \beq\label{ptraj}
 x\,\equiv\,x^1\,=\,\upsilon t\,,\quad x^2\,=\,x^3\,=\,0\,,\quad
 z\,=\,\frac{1}{\gamma}\,t,\quad x^2+z^2\,=\,t^2\,,\eeq
where $\gamma=1/\sqrt{1-\upsilon^2}$ is the boost factor associated with
the longitudinal motion along the boundary. For the examples considered
in Sects.~\ref{SmallPulse} and \ref{TLWP} we actually need\footnote{Note
that when comparing the present discussion to that in
Sect.~\ref{SmallPulse}, the `free particle' which is currently under
consideration corresponds to the small pulse falling down the string, and
{\em not} to the heavy quark that the string is attached to.} $\upsilon
=0$, but for the present discussion it is not more difficult to consider
a general value $\upsilon<1$. Note that \eqnum{ptraj} together with the
fact that $z\ge 0$ restricts the time variable to positive values, $t\ge
0$. That is, here we do not consider a {\em steady} situation, which
would be translationally invariant in time, but a situation where the
particle is put at $t=0$ on the Minkowski boundary and left to freely
fall into AdS$_5$.

To construct the associated bulk energy--momentum tensor, it is
convenient to start with the expression valid for a massive particle with
mass $m$, and then take the limit $m \to 0$. Then the 5D stress tensor
reads
 \beq\label{tmn particle}
 t^{MN} = \frac{m}{\sqrt{-g}}\,
 \frac{\dif x^M}{\dif t}\,
 \frac{\dif x^N}{\dif t}\,
 \frac{\dif t}{\dif \tau}\,
 \delta^{(4)}(x - x_p(t)),
 \eeq
with $\tau$ the proper time and $x_p(t)$ the particle trajectory. We have
 \beq
 \dif \tau = \sqrt{-g_{00}}\,\frac{\dif t}{\gamma}.
 \eeq
Furthermore, the momentum in local inertial coordinates can be expressed
in terms of the conserved momentum as 
 \beq
 m \gamma = \frac{E_0}{\sqrt{-{g_{00}}}},
 \eeq
so that \eqnum{tmn particle} becomes
 \beq
 t^{MN} = \frac{E_0}{|g_{00}|\sqrt{-g}}\,
 \frac{\dif x^M}{\dif t}\,
 \frac{\dif x^N}{\dif t}\,
 \delta^{(4)}(x - x_p(t)).
 \eeq
At this level we can take the massless limit and specialize to AdS$_5$
and the trajectory in \eqnum{ptraj}. One immediately finds (below,
$X^M=(t,\upsilon t,0,0,t/\gamma)$)
 \beq\label{tmnpv}
 t^{MN}\,=\,E_0
 \left(\frac{z}{L}\right)^7\delta(x-\upsilon t)
\, \delta^{(2)}(\bxT) \delta(z-t/\gamma)\,\frac{X^M X^N}{t^2}\,.
 \eeq
After lowering the indices and letting $\upsilon=0$, this yields
 \beq\label{tmnpart}
 t_{00} = t_{55} = -t_{05}\, =\,
 E_0 \left(\frac{z}{L}\right)^3
 \delta^{(3)}(\br) \delta(z-t), \qquad t_{iM} =0,
 \eeq
which is very similar to the stress tensor \eqref{tradPulse} for the
radiation emitted by a small string perturbation. The only, inessential,
difference between Eqs.~\eqref{tradPulse} and \eqref{tmnpart} is that, in
the latter, the bulk energy density is a strict $\delta$--function at
$z=t$, while in the former it has some width around the maximum at $z=t$.
Thus, the fact that the source in the bulk propagates at the speed of
light along the radial direction ($\upsilon_z=1$) automatically implies
the tensor structure in \eqref{tradPulse} or \eqref{tmnpart} for the
associated bulk stress tensor.

Using the general formul\ae{} in Sect.~\ref{Constant}, it is
straightforward to compute the boundary energy density produced via
backreaction from the bulk stress tensor \eqref{tmnpv}. One thus finds
that $\EB=0$. Furthermore, within the integrand of \eqnum{EA} we can
write
 \beq\hspace*{-.6cm}
z (2 t_{00} - t_{55}) - (t-\tp) t_{05} + (x - \xp)^i t_{i5}
 =E_0 \left(\frac{z}{L}\right)^3
 \frac{t+vx}{\gamma}\,\delta(\xp-\upsilon \tp)
 \, \delta^{(2)}(\acute{\bm{x}}_{\perp}) \delta(z-\tp/\gamma).\nonumber
 \eeq
Using the $\delta$--functions above to perform the integrations over
$\xp$, $\acute{\bm{x}}_{\perp}$ and $z$, one obtains
 \beq\label{EApart}
 \mcal{E}\,=\,\mcal{E}_A\,=
 \frac{2E_0}{\pi}\,\frac{t+vx}{\gamma^2}\
  \del^2_{r^2} \int_0^\infty \dif \tp\, \tp \
  \delta\big(t^2 -r^2 - 2(t-\upsilon x)\tp\big)\,
 \,.\eeq
Note that the $\delta$--function inside the integrand together with the
lower limit on $\tp$  (which was generated by the condition $z\ge 0$)
imply a factor $\Theta(t^2-r^2)$. This is the expected condition,
following from causality, that the space--time distribution of the energy
density on the boundary must be time--like. However, this
$\Theta$--function becomes a $\delta$--function after performing the two
external derivatives w.r.t. $r^2$. One has indeed
 \beq\label{ed particle}
 \mcal{E} \,=\, \frac{E_0}{2 \pi\gamma^2}\,\frac{t+vx}{(t-\upsilon x)^2}\
  \del^2_{r^2} [(t^2 - r ^2) \Theta(t^2-r^2)]\,=\,
 \frac{E_0}{2 \pi\gamma^2}\,\frac{t+vx}{(t-\upsilon x)^2}\,\delta(t^2-r^2)
  \,.
 \eeq
\eqnum{ed particle} describes a spherical shell of zero width which
propagates at the speed of light, so like classical radiation, and whose
shape looks anisotropic because of the boost with velocity $\upsilon$ in
the $x$ direction. By taking $\upsilon\to 0$, isotropy becomes manifest:
 \beq\label{shell} \mcal{E}\, = \,\frac{E_0}{4 \pi r^2}\,
 \delta(t-r)\,,\qquad \frac{\rmd E}{\rmd\Omega}\,\equiv\int \rmd r \,r^2
 \mcal{E}\,=\,\frac{E_0}{4 \pi}\,.
 \eeq
In the above equation we have also shown the energy density per unit
solid angle. Clearly, the total energy is equal to $E_0$ and coincides,
as expected, with the total energy stored in the bulk stress tensor
\eqref{tmnpart}.

Returning to the calculation of the energy density according to
\eqnum{EApart}, we note that the condition $t=r$ implies $\tp=z=0$. That
is, the whole backreaction on the boundary comes from the string endpoint
at $z=0$, so like for the string perturbation in Sect.~\ref{SmallPulse}.
In both cases, this property follows from the fact that the source in the
bulk --- the pulse on the string, or the falling particle --- propagates
at the 5D speed of light.

To conclude this discussion, let us mention that for the bulk excitation
under discussion --- a free, point--like, massless particle with the
stress tensor \eqref{tmnpv} --- the backreaction on the AdS$_5$ metric
can be computed {\em exactly}, and not only in the linear approximation
in which hold our general formul\ae{} \eqref{EA}--\eqref{EB} for $\EA$
and $\EB$. The corresponding solution is a gravitational shock--wave in
AdS$_5$, whose intersection with the boundary is the spherical shell in
\eqnum{ed particle}. At generic points inside AdS$_5$, the gravitational
shock--wave involves the $\delta$--function $\delta(t^2-r^2-z^2)$. This
will be further discussed in Sect.~\ref{sect-flucts}.

\subsection{A time--like wave--packet}
\label{TLWP}

In this subsection, we shall consider a problem which is perhaps better
motivated at a physical level than the problem of the falling particle
discussed in the previous subsection, but whose mathematical treatment
turns out to be very similar: the decay of a time--like wave--packet.
This wave--packet, which acts as a boundary condition for a supergravity
field in the bulk, is to be seen as a model for the virtual photon
created via $e^+e^-$ annihilation. Strictly speaking, a photon within
${\mathcal N}=4$ SYM should be described by a vector field coupled to the
${\mathcal R}$--current operator, but for the present purposes there is
no loss of generality if we instead consider a scalar field coupled to
the `glueball' operator ${\rm Tr}\,F^2_{\mu\nu}$.

Working in the rest frame of the virtual `photon', we are led to consider
the following boundary condition for the associated dilaton field in the
bulk:
 \beq\label{phib}
 \phi_b = \exp\left(-\rmi \omega t - \frac{t^2}{2 \sigma^2}
 - \frac{r^2}{2 \sigma_r^2}\right).
 \eeq
Clearly, this is a wave--packet with zero average momentum\footnote{The
overall normalization of the  wave--packet will be fixed later on.}. To
also guarantee that this is time--like (TL), we need to assume that both
$\sigma$ and $\sigma_r$ are relatively large: $\sigma\omega\gg 1$ and
$\sigma_r\omega\gg 1$. Then the Fourier modes introduced by the Gaussian
in \eqnum{phib} are small compared to $\omega$, which therefore sets the
virtuality of the wave--packet.

Before we solve the AdS problem with boundary condition \eqref{phib}, let
us first study the corresponding classical problem. In that case, the
wave--packet \eqref{phib} should be viewed as a source in the r.h.s. of
the massless Klein--Gordon equation in the usual, Minkowski, space--time.
The respective retarded solution reads
 \beq\label{rho to phi}
 \phi(t,\br) = \int \dif \tp\,\dif^3\brp\,
 \Theta(t - \tp)\,
 \frac{\delta(\tp -t + |\br - \brp|)}{|\br - \brp|}\,
 \rho(\tp,\acute{{\br}}),
 \eeq
where $\rho(\tp,\acute{{\br}})$ is the same function as in \eqnum{phib},
but now viewed as a classical source. We are interested in the response
at large distances $r \gg \sigma,\sigma_r$, and therefore  $\rp \ll r$
(indeed, the Gaussian in the source sets effectively $\rp \lesssim
\sigma_r$). In this regime we can approximate (with $\alpha$ the angle
between $\br$ and $\brp$)
 \beq
 |\br - \brp| \simeq r - \rp \cos\alpha + \frac{\rp^2}{2 r}\,
 \sin^2 \alpha\,.
 \eeq
By keeping only the first two terms in this expansion, it is
straightforward to find
 \beq\label{phiclass}
 \phi \simeq \frac{1}{r}\,
 \exp\left[-\rmi \omega (t-r) - \frac{(t-r)^2}{2 \sigma^2} -
 \frac{\sigma_r^2 \omega^2}{2}\right].
 \eeq
Since $\sigma_r\omega\gg 1$ by assumption, it is clear that this field
$\phi$ is strongly suppressed (in fact, truly negligible) in the regime
$r \gg \sigma_r$ where the above approximation is valid. This leads us to
the expected conclusion that, being off--shell, a classical TL
wave--packet cannot propagate outside its original support at $\rp
\lesssim \sigma_r$.

We now turn to the AdS calculation at strong coupling, where \eqnum{phib}
acts as the 4D boundary limit for the solution to the 5D Klein--Gordon
equation in AdS$_5$. Then the solution can be expressed with the help of
the respective boundary to bulk propagator:
 \beq\label{phib to phi}
 \phi(t,\br) = \int \dif \tp\,\dif^3\brp\,
 \mcal{D}(t-\tp,\br - \brp,z)
 \phi_b(\tp,\acute{{\br}}).
 \eeq
A convenient and rather compact expression for this propagator in
coordinate space, to be derived in Appendix \ref{appProp}, is
 \beq
 \mcal{D}(t-\tp,\br - \brp,z) = -
 \frac{2 z^4}{\pi}\,\Theta(t- \tp) \delta'''(\mcal{W}),
 \eeq
with $\mcal{W}$ defined in \eqnum{W}. Then we can write \eqnum{phib to
phi} as
 \beq
 \phi = - \frac{z^4}{\pi}\,\del^3_{z^2}
 \int \dif \tp\,\dif^3\brp\,
 \frac{\delta[\tp - t + \sqrt{z^2 + (\br - \brp)^2}]}
 {\sqrt{z^2 + (\br -\brp)^2}}\,\phi_b(\tp,\brp).
 \eeq
Once again we consider large distances $r \gg \sigma,\sigma_r$, where we
can expand
 \beq
 \sqrt{z^2+(\br - \brp)^2} \simeq \rho - \frac{r \rp}{\rho}\, \cos\alpha + \frac{r^2 \rp^2}{2 \rho^3}\, \sin^2 \alpha,
 \eeq
with $\rho = \sqrt{z^2+r^2}$, and we keep the most dominant terms in the
exponential. We first integrate over $\alpha$ and subsequently over
$\brp$, and finally we take the three derivatives on the factor which
varies the most, that is $\exp(\rmi \omega \rho)$, to find
 \beq\label{phiz}
 \phi = \frac{\rmi \sqrt{2\pi}\sigma_r^3 \omega^3}{4}\,
 \frac{z^4}{\rho^4} \exp\left[-\rmi \omega
 (t-\rho) - \frac{(t-\rho)^2}{2 \sigma^2} -
 \frac{\sigma_r^2 \omega^2 r^2}{2 \rho^2}\right].
 \eeq

It is instructive to compare this bulk field at strong coupling to the
classical field in \eqnum{phiclass}: the last term in the argument of the
Gaussian in \eqref{phiz} is now proportional to $r^2/\rho^2$, and hence
it can remain small even for $\sigma_r\omega\gg 1$ provided $r$ is
sufficiently small compared to $\rho$, meaning $r\ll z$. In other terms,
the wave--packet can propagate within AdS$_5$, but only within a small
angle $\theta \equiv r/z\lesssim 1/(\sigma_r \omega)$ with respect to the
radial axis. When $\sigma_r\omega\gg 1$ this angle is so small that we
can neglect the displacement along the spatial direction $r$ and
approximate $\rho \simeq z$. Then \eqnum{phiz} becomes
 \beq\label{phiz2}
 \phi \,= \,c \exp\left[-\rmi \omega (t-z) -
  \frac{(t-z)^2}{2 \sigma^2} - \frac{r^2}{2 \theta^2 z^2}\right]\,,
 \eeq
where $c$ is a dimensionless constant which depends on the various energy
scales of our problem, but not upon the coordinates (since we have
replaced $z/\rho=1$). At this level, it is convenient to view $c$ as a
free constant, to be later fixed by the condition that the total energy
carried by the TL wave--packet takes some prescribed value $E_0$.

In order to compute the backreaction due to this bulk wave--packet, we
need the associated stress tensor. For a scalar field in AdS$_5$ space,
this is given by
 \beq
 t_{MN} = \frac{N_c^2}{16 \pi^2 L^3}
 \left[\del_{M}\phi\, \del_{N}\phi^* +
 \del_{N}\phi \,\del_{M}\phi^*
 -g_{MN}\,g^{PQ}\, \del_P\phi\, \del_Q\phi^*\right].
 \eeq
Under the present assumptions, we have $z \gg r\gg \sigma$. Using these
conditions to simplify the components of $t_{MN}$, we eventually find
 \beq\label{tmn wave--packet}
 t_{00} = t_{55} = -t_{05} = \frac{N_c^2 c^2 \omega^2}{16 \pi^2 L^3}
 \exp\left[- \frac{(t-z)^2}{\sigma^2} - \frac{r^2}{\theta^2 z^2}\right],
 \qquad t_{iM} = 0.
 \eeq
In order to arrive at these results we have dropped, in all components of
$t_{MN}$, terms which are suppressed, compared to the leading
non-vanishing components, by powers of $1/\sigma \omega$ and $\sigma/z$.
Recalling that $\theta\to 0$, we can replace in the above
  \beq\label{smallr}
 \exp\left(- \frac{r^2}{\theta^2 z^2}\right)
 \simeq \pi^{3/2} \theta^3 z^3 \delta^{(3)}(\br)\,.
 \eeq
Since on the other hand we have assumed $r\gg \sigma$ in order to arrive
at \eqnum{tmn wave--packet}, it is clear that the use of the above
$\delta$--function is legitimate only when looking at the bulk stress
tensor over spatial distances much larger than $\sigma$. Remarkably, the
(approximate) mathematical identity in \eqnum{smallr} introduces a factor
$z^3$ in $t_{00}$, which from the manipulations in
Sects.~\ref{SmallPulse} and \ref{Falling} we know to be essential in
order to obtain a result for the energy density which gets contributions
only from the boundary at $z=0$.


With \eqnum{smallr}, our bulk tensor takes the form
 \beq\label{t00 wave--packet}
 t_{00} = t_{55} = -t_{05} = E_0 \,\frac{z^3}{L^3}\, \frac{1}{\sqrt{\pi}\sigma} \,
 \exp\left[-\frac{(t-z)^2}{\sigma^2}\right]\,\delta^{(3)}(\br)
 \eeq
where we have chosen the constant $c$ so that
 \beq
 c^2 = \frac{16 E_0}{N_c^2 \sigma \omega^2 \theta^3} =
 \frac{16 E_0 \omega \sigma_r^3}{N_c^2 \sigma}\,.
 \eeq
It is of course understood that \eqnum{t00 wave--packet} can be used for
computing the backreaction only at sufficiently large distances $r\gg
\sigma$.

As expected, \eqnum{t00 wave--packet} is very similar to the respective
expressions for the small pulse down the string, \eqnum{tradPulse}, and
for the falling massless particle, \eqnum{tmnpart}. Once again, we have
the bulk stress tensor of a source propagating along the radial direction
of AdS$_5$ at the speed of light. From our previous experience, we expect
the corresponding boundary energy to be generated exclusively by the bulk
field at $z=0$ and to be localized near $r=t$ within a distance $\sigma$.
We have indeed $\EB=0$ and (compare to \eqnum{EintPulse})
 \beq
 \mcal{E} = \EA = \frac{E_0}{\pi^{3/2} \sigma}\, \del^2_{r^2}
 \int_{0}^{\infty} \dif z\,
 \frac{z (z+ \sqrt{z^2 + r^2})}{\sqrt{z^2 + r^2}}\,
 \exp\left[-\frac{(t - z- \sqrt{z^2+r^2})^2}{\sigma^2} \right],
 \eeq
which after manipulations entirely similar to those already encountered
in Sect.~\ref{SmallPulse} (involving in particular the change of variable
\eqref{xi}) finally yields the following result for the energy density
produced by the decay of the TL wave--packet:
 \beq\label{ed wave--packet}
 \mcal{E} = \frac{E_0}{4 \pi r^2}\,\frac{1}{\sqrt{\pi}\sigma}\,
 \exp\left[-\frac{(t - r)^2}{\sigma^2} \right]\,.
 \eeq
As anticipated this is a narrow spherical shell propagating at the speed
of light, with the width set by the initial perturbation. Moreover, this
energy density has been entirely produced via backreaction from the
endpoint of the integration at $z=0$.

Whereas the result \eqref{ed wave--packet} shows the same type of
space--time pattern as the radiation by a small string perturbation, cf.
\eqnum{EPulse}, and it has been generated via similar mathematical
manipulations, there is nevertheless an important difference between the
two: the total energy in \eqnum{ed wave--packet} is independent of
$\lambda$ and thus it remains constant in the strong coupling limit. This
is as expected on physical grounds, since the original `virtual photon'
has only a finite energy $E_0$. But this raises an uncomfortable
relationship between this result, \eqnum{ed wave--packet}, and a quantum
mechanical interpretation of the outgoing radiation. Namely, it seems
difficult to understand how it is possible to distribute a finite amount
of energy over a thin spherical shell which expands for ever (so that the
energy density can become arbitrarily small) while keeping a constant,
narrow, width $\sigma$ (so that the radial wave number $k_r$ cannot
decrease below a value $k_0\simeq 1/\sigma$ fixed by the uncertainty
principle). Indeed, this would imply that a finite--size detector which
is located sufficiently far away can register a non--zero amount of
energy which is smaller than $\hbar k_0$, that is, smaller than the
energy of a single quanta with that wave number. Classically, such a
situation would be permitted, since the amplitude of the classical field
corresponding to wave number $k_r\simeq k_0$ can decrease with $r$ and
become arbitrarily small. But in quantum mechanics, the amplitude squared
is proportional to the number of quanta per mode, which takes only
discrete values.

\section{String fluctuations and the freely falling particle}
\label{sect-flucts}

In the previous sections, we have analyzed various perturbations in the
bulk associated with radiating sources on the boundary and shown that
several of these perturbations (a small pulse flowing down along a string
or a dilaton, time--like, wave--packet) can be mathematically idealized
as a massless point particle falling freely into the radial dimension $z$
of AdS$_5$. For all these cases, we have found that the energy flow on
the boundary ({\em i.e.}, in the physical space), as calculated from the
backreaction of the 5D energy--momentum tensor associated with the bulk
perturbation, is a thin spherical shell of matter moving outwards at the
velocity of light.

Going beyond the supergravity approximation, all the bulk perturbations
that were previously described by local supergravity fields should be
replaced by microscopic strings. Similarly, for the problems involving a
macroscopic string one should add the effects of stringy excitations. In
this section, we would like to show that the world--sheet fluctuations of
the microscopic strings can have sizeable effects on our previous
results. To that aim, we shall take as an example the closed string
underlying the `falling point particle' introduced in
Sect.~\ref{Falling}. As we shall see, depending on what measurement is
made, the `point--like particle' may be a good approximation, with string
fluctuations unimportant, or a poor approximation, having important
string fluctuations.

Hofmann and Maldacena  \cite{Hofman:2008ar} have already studied string
fluctuations for the falling particle but only under the circumstance
where energy at a given angle, or set of angles, is measured integrated
over the three--dimensional radial coordinate. They found string
corrections to angular energy--energy correlations to be small and
proportional to $1/\lambda$. However, the purpose of our study is to see
whether the out--flowing energy remains in a thin radial shell when
string fluctuations are included, and to carry out that calculation we
need to generalize the discussion of \cite{Hofman:2008ar}.

\subsection{A special set of light--cone coordinates}
\label{sect-LC}

In order to describe string fluctuations, it is useful to use light--cone
coordinates where the free string in flat space can be easily quantized
\cite{Polchinski:1998rq,Zwiebach:2004tj}. Of course we are not in flat
space, but so long as the fluctuations are not too large and we use
coordinates where the string does not move in the fifth dimension, a flat
space calculation should be adequate, at least for qualitative estimates.
Clearly, our present coordinates, cf. \eqnum{metric}, are not well suited
in that sense, since in these coordinates the `point particle' is moving
along the $z$ direction. A strategy to circumvent this problem, that was
also followed in Ref.~\cite{Hofman:2008ar}, is to make a change of
coordinates which mixes radial with boundary coordinates, in such a way
that the bulk motion of our `falling particle' occurs at a fixed value of
the {\em new} fifth dimension. A suitable set of coordinates in that
sense has been originally introduced in \cite{Cornalba:2007fs} and later
applied to $e^+e^-$ annihilation in \cite{Hofman:2008ar,Hatta:2008st}.

We shall connect these new coordinates to our previous ones by using the
six global coordinates $W_M$ on AdS$_5$, which satisfy
\beq
W_{-1}^2 + W_0^2 -W_1^2 -W_2^2 - W_3^2 -W_4^2 = L^2\,.
\eeq
These are related to our previous coordinates in \eqnum{metric} via
 \beq
 W_{-1}+W_4= \frac{L^2}{z}\,, \quad W_{\mu}=\frac{x^\mu L}{z}\,,
 \quad (\mu=0,1,2,3)\,.
 \eeq
In terms of them, our new coordinates, $y^M$, are defined as
 \beq
 W_0+W_3 = \frac{L}{y_5}\,,
  \quad W_{-1} = -\frac{y^0L}{y_5}\,, \quad
   W_4 = -\frac{y^3L}{y_5}\,, \quad  W_{1,2}=\frac{y^{1,2}L}{y_5}\,,
 \eeq
and hence they are dimensionless. In these variables, the metric
\eqref{metric} becomes
\beq
\rmd s^2 \,=\, \frac{L^2}{y_5^2} \left[-2\rmd y^+ \rmd y^- + \rmd
\bm{y}_\perp^2+\rmd y_5^2\right]\,,
\eeq
with $y^\pm =(y^0\pm y^3)/{\sqrt{2}}$ and $\bm{y}_\perp=(y^1,y^2)$. The
following relations hold between the $x$ and $y$ variables (below $x^\pm
= (x^0\pm x^3)/{\sqrt{2}}$ and $\bm{x}_\perp=(x^1,x^2)$)
\beq
y^+ = -\frac{L}{2x^+}\,, \quad y^- = \frac{2x^+x^- - \bm{x}_\perp^2
-z^2}{2x^+L}\,, \quad  \bm{y}_\perp = \frac{\bm{x}_\perp}{\sqrt{2}x^+}\,,
\quad y_5= \frac{z}{\sqrt{2}x^+}\,. \label{cornalba}
\eeq
One also has $y_5= -{\sqrt{2}y^+ z}/{L}$. In particular, the physical
boundary at $z=0$ lies at $y_5=0$ in these new coordinates.

Using \eqnum{cornalba} one can now easily check that these new variables
realize indeed the desiderata for which they have been introduced: with
respect to them, the bulk motion of the `point particle' introduced in
Sect.~\ref{Falling} takes place at a fixed value of the (new) radial
coordinate $y_5$. Indeed, we can write (compare to \eqnum{tmnpart})
 \beq
 z^3 L\, \delta^{(3)}(\br) \delta(z-t)\,=\,\sqrt{2}\,\delta(y_5-1)
 \delta^{(2)}({\bm{y}_\perp}) \delta(y^-)\,.\eeq
In these variables, the bulk excitation looks like a point--like particle
propagating along the $y^3$ direction at the speed of light ($y^3=y^0$,
or $y^-=0$), while keeping a fixed value $y_5=1$ away from the boundary.
It is easy to check that the only non--zero component of the particle
stress tensor, \eqnum{tmnpart}, in these new coordinates is\footnote{For
more clarity, we shall systematically indicate tensor components in the
$y$ variables by a hat. Note also that, if one uses the coordinates $y^0$
and $y^3$ instead of the light--cone coordinates $y^\pm$, then the
non--zero components of $\hat t_{MN}$ are $\hat t_{00}=\hat t_{33}=-\hat
t_{03}$, with $\hat t_{--}=2\hat t_{00}$.}
 \beq\label{hatt}
 \hat t_{--}\,=\,\frac{E_0}{\sqrt{2}L^2}\,\delta(y_5-1)
 \delta^{(2)}({\bm{y}_\perp}) \delta(y^-)\,.\eeq
Note that the spherical symmetry of the problem, which was manifest in
our original spatial coordinates $\br=(x^1,x^2,x^3)$, appears now to be
lost. But this is not a serious drawback, as the full symmetry will
reappear when transforming the final results for the energy density back
to the original coordinates.

The new variables introduced above have another virtue, which will
greatly simplify our subsequent discussion of the string fluctuations:
the backreaction associated with the stress tensor \eqref{hatt} is
particularly simple to compute. Indeed, with this source, the 5D system
of Einstein equations reduce to a single, linear, equation for $\delta
\hat g_{--}$ which can be solved exactly, using the bulk--to--bulk
propagator in AdS$_5$. The solution is a gravitational shock--wave which
at any point in the bulk is proportional to $\delta(y^-)$ (see e.g.
\cite{Gubser:2008pc,Gubser:2009sx,Hatta:2008st}). Then the
energy--momentum tensor on the boundary is extracted in the standard way,
from the behavior of $\delta \hat g_{--}$ near $y_5=0$. One thus finds
that the boundary response is also a shock--wave, with (below
$p^+=E_0/\sqrt{2}$)
 \beq\label{hattbound}
 \hat T_{--}(y^+,y^-, \bm{y}_\perp)
 \,=\,\frac{2}{\pi}\,\frac{p^+ L}{(1+\bm{y}_\perp^2)^3}
 \, \,\delta(y^-)\,.\eeq
Recalling that on the boundary ($y_5=z=0$) one has $y^-\propto 2x^+x^- -
\bm{x}_\perp^2=t^2-\br^2$, it becomes clear that what looks like a
shock--wave ($y^-=0$) in the $y$--coordinates is in fact the spherical
shell propagating at the speed of light ($t=r$) that we have previously
found in Sect.~\ref{Falling}, cf. \eqnum{shell}.

This is a quite remarkable fact:  the property of lack of broadening that
we observed when computing the radiation in the original coordinates
appears as {\em Lorentz contraction} in these new coordinates. Similarly,
our observation in Sect.~\ref{Falling}, that the radiation by a massless
particle falling into AdS$_5$ comes only from $\tp=z\to 0$ can be viewed
as a consequence of {\em Lorentz time dilation} with respect to the
`time' $y^0$ (or $y^+$). Indeed, given the discussion towards the end of
Sect.~\ref{Constant}, it should be clear that the shock--wave in
\eqnum{hattbound} has been fully generated via radiation at early times
$\acute y^+\to -\infty$. Returning to the original variables, this
implies $\acute x^+\to 0$ and therefore $\tp=z\to 0$ (recall that $\acute
x^3=0$ for this source). Since the propagation at the speed of light is
of course essential for such `kinematical' arguments using the $y$
coordinates, it becomes obvious that the absence of backreaction from
sources at large values of the `original' 5th coordinate $z$ is a
consequence of the fact that these sources propagate in AdS$_5$ at the
speed of light.

Given a result like \eqref{hattbound} for the (LC) energy density $\hat
T_{--}$ in the $y$ coordinates, it is useful to know how to transform it
back to the original $x$ coordinates. This requires a change of variables
$y^\mu\to x^\mu$ together with a Weyl transformation (see
Refs.~\cite{Hofman:2008ar,Hatta:2008st} for details). We here present the
result of this transformation for the case of the energy density per unit
solid angle. In the original coordinates, this is defined as
\beq
\frac{\rmd E}{\rmd \Omega}(\bm{n})\,\equiv
 \,\lim_{t\to \infty}\int \rmd r \,r^2\,
 \mcal{E}(t,\br)\,=\,\lim_{r\to \infty} \,r^2\int^\infty_0 \rmd t\, n_i
T^{0i}(t,r\bm{n})\,, \label{energyop}
\eeq
where the unit vector $\bm{n}=\bm{r}/r$  specifies the direction of
measurement and the second equality (which is the formula used in the
analysis in Ref.~\cite{Hofman:2008ar}) exploits energy--momentum
conservation ($\del^\mu T_{0\mu}=0$) to express ${\rmd E}/{\rmd \Omega}$
in terms of the energy flux along $\bm{n}$ integrated over all times and
evaluated at points on the sphere at $r\to\infty$. It should be clear
from above formul\ae{} that ${\rmd E}/{\rmd \Omega}$ contains no
information about the {\em radial} distribution of the energy flow: this
is averaged out when performing the integration over $r$, or over time.


For the case of a particle falling in AdS$_5$, one can express either
$\mcal{E} \equiv T_{00}$ or the Poynting vector $S^i=T^{0i}$ in terms of
$\hat T_{--}$, and thus obtain \cite{Hofman:2008ar,Hatta:2008st}
 \beq
\frac{\rmd E}{\rmd \Omega}(\bm{n})\,=\,
\frac{\sqrt{2}(1+\bm{y}^2_\perp)^3}{8L} \int \mbox{d}y^- \,\hat
T_{--}(y^+=0,y^-,\bm{y}_\perp)\,. \label{flat}
\eeq
In the above equation, the condition $y^+=0$ expresses the limit
$r\to\infty$ (and hence $x^+\to\infty$). Also, $\bm{y}_\perp$ is related
to $\bm{n}=(\sin \theta \cos \phi, \sin \theta \sin \phi, \cos \theta)$
via
 \beq\label{yperpn}
 \bm{y}_\perp = \frac{\sin \theta}{1+\cos \theta}\,(\cos \phi, \sin \phi)\
 \Longrightarrow \  \frac{2}{1+\bm{y}_\perp^2}=1+\cos \theta\,.
 \eeq
(This follows from $\bm{y}_\perp=\bm{x}_\perp/(t+x^3)$ after using $t=r$,
as appropriate for the spherical shell in \eqnum{hattbound}.) Using the
above relations together with \eqnum{hattbound} for $\hat T_{--}$ it is
straightforward to check that ${\rmd E}/{\rmd \Omega} = E_0/4\pi$, in
agreement with the respective result in \eqnum{shell}.

\subsection{Longitudinal string fluctuations and broadening}
\label{Longit}

As shown by Hofman and Maldacena \cite{Hofman:2008ar} the $y$ coordinates
introduced in the previous subsection allow for a heuristic treatment of
the world--sheet fluctuations of a microscopic string falling into
AdS$_5$ which parallels the corresponding analysis in flat space and in
the light--cone gauge (see e.g.
\cite{Polchinski:1998rq,Zwiebach:2004tj}). In what follows we shall
perform a similar analysis but focus on a different problem as compared
to Ref.~\cite{Hofman:2008ar}. Namely, we shall focus on the {\em
longitudinal} string fluctuations $\delta y^-(\tau,\sigma)$, which play
no role in the calculation of the {\em angular} distribution of the
energy --- as obvious from Eq.~\eqref{flat}, which involves an integral
over all values of $y^-$ --- but which have the essential effect to
broaden its {\em radial} distribution (in $r=|\br|$), as we shall see.


Our strategy to include string fluctuations will be as follows. As
previously discussed, in the absence of fluctuations the closed string
reduces to a pointlike, massless, particle that in the $y$ representation
is located at $y_5=1$, $\bm{y}_\perp=0$ and $y^-=0$, cf. \eqnum{hatt}.
The effect of the string fluctuations is to render this distribution
`fuzzy'~: a fluctuation $\delta y^M(\tau,\sigma)$ in the string
worldsheet is responsible for a contribution to $\hat t_{MN}$ with
support at $y_5=1+\delta y_5$, $y^i=\delta y^i$ (with $i=1,2$) and
$y^-=\delta y^-$. There is no fluctuation in $y^+$ by construction,
because we shall work in the light--cone (LC) gauge
 \beq\label{LCdef} y^+=\frac{\alpha'p^+ \tau}{L}\,,
 \eeq
with $p^+=E_0/\sqrt{2}$. Some general results on the quantization of
string fluctuations in LC gauge and flat space are reviewed in Appendix
\ref{SLC}. (The discussion in Appendix \ref{SLC} is heuristically
extended here to AdS$_5$, with $y_5$ treated as one of the `transverse
coordinates', on the same footing as $y^1$ and $y^2$.) As we recall in
that Appendix, $y^-(\tau,\sigma)$ is a dependent variable whose
fluctuations come from those in $\bm{y}_\perp$ and $y_5$ (see
\eqnum{constraint}). However, a given string mode only contributes
$\delta y_5\sim \delta y^i \sim 1/\lambda^{1/4}$ while it gives $\delta
y^-\sim 1$ (see Eqs.~\eqref{dyi} and \eqref{dy-}). Moreover, the
equal--point correlations\footnote{Throughout this section, the brackets
$\langle \cdots\rangle$ refer to the average over the string
fluctuations.} $\langle \delta y_5^2\rangle$ and $\langle (\delta
y^i)^2\rangle$ have only weak, logarithmic, ultraviolet divergences in
flat space (see \eqnum{suss}) and have been argued to be finite in
AdS$_5$ \cite{Polchinski:2001ju}. By contrast, $\langle (\delta y^-)^2
\rangle$ shows a strong, quadratic, UV divergence in flat space cf.
\eqnum{fluct}, thus yielding a potentially large contribution even after
the introduction of an {\em ad--hoc} cutoff on the number of modes.
Besides, we do not know whether one can generalize the arguments in
\cite{Polchinski:2001ju} in such a way as to also cure this divergence
when moving from flat space to AdS$_5$. We shall return later to these
points.

For these reasons, we henceforth neglect the fluctuations in $y^i$ and
$y_5$ and concentrate on those in $y^-$. (The effects of $\delta y^i$ on
the angular distribution of the energy have been estimated in
Ref.~\cite{Hofman:2008ar} and found to be small, of
$\order{1/{\lambda}}$.) Also, since $y^-=0$ for the classical pointlike
particle in the bulk, cf. \eqnum{hatt}, we denote the respective
fluctuations simply as $\acute y^-(\tau,\sigma)$. As usual, the acute
symbol in $\acute y^-$ is used to differentiate a point on the string
from the point $y^-$ on the boundary at which we measure the LC energy
density $\hat T_{--}$. Finally, in order to be able to measure the effect
of the fluctuations, we need to give up the integral over $y^-$ in
\eqnum{flat}, but rather study the $y^-$--dependence of the LC energy
density $\langle\hat T_{--}\rangle$ (averaged over the string
fluctuations).

To summarize, we need to compute the backreaction from a string
fluctuation with longitudinal coordinate $\acute y^-(\tau,\sigma)$ and
then average over the fluctuations, according to the rules in Appendix
\ref{SLC} heuristically extended to AdS$_5$. As already mentioned in the
previous subsection, one advantage of the $y$ representation is that the
backreaction due to the stress tensor \eqnum{hatt} is easy to compute.
Specifically, the energy density $\hat T_{--}$ on the boundary is
obtained as the convolution of the bulk point--like source with the
bulk--to--boundary propagator (that we here take from
Ref.~\cite{Hofman:2008ar}). This yields
\beq\label{ali}\hspace*{-0.8cm}
\langle \hat T_{--}(y^-,\bm{y}_\perp)\rangle  &=& \frac{6\rmi \,p^+
L}{\pi^2}  \int_{-\infty}^\infty \rmd \acute y^+
 \int_0^{2\pi} \frac{\rmd\sigma}{2\pi} \\
 &\times& \left\langle
\frac{\acute y_5^4(\tau, \sigma)}{[-2\acute y ^+(\acute y ^-(\tau,\sigma)
- y^-)+ \acute y_5^{2}(\tau,\sigma)+
(\bm{y}_\perp-\acute{\bm{y}}_\perp(\tau,\sigma))^2+
i\epsilon]^4}\right\rangle\,, \nonumber
\eeq
where we shall immediately take $\acute y_5=1$ and
$\acute{\bm{y}}_\perp=0$, since we neglect the respective string
fluctuations, as already explained. Also, it is here understood that
$y^+=0$, corresponding to $x^+\to\infty$. At this stage it is convenient
to insert unity in the form
\beq
1=\int \rmd \dacute y ^- \delta(\dacute y ^- - \acute y
^-(\tau,\sigma))\,,
\eeq
and thus get
\beq\label{limit}
\langle \hat T_{--}(y^-,\bm{y}_\perp)\rangle  &=& \frac{6\rmi
p^+L}{\pi^2}\int \frac{ \rmd \acute y ^+ \rmd \dacute y ^-}{[-2\acute y
^+(\dacute y ^- - y^-)+ 1+\bm{y}_\perp^2+i\epsilon]^4} \,P(\dacute y ^-)
\,,
\eeq
where
\beq\label{Pdef}
P(\dacute y ^-)\equiv \int\frac{\rmd \sigma}{2\pi} \left\langle \delta
(\dacute y ^- - \acute y ^-(\tau,\sigma))\right\rangle\,.
\eeq
is recognized as the probability distribution for a point on the string
to be found at $\dacute y ^-$. In particular, it satisfies
\beq\label{prob1}
\int \rmd \dacute y^- P(\dacute y ^-)=1\,,
\eeq
as it should. Because of translation invariance we expect $P$ to be
independent of $\tau$. So, the only dependence upon $\acute y ^+$ in the
integrand of \eqnum{limit} is that explicit in the denominator. It is
then straightforward to perform the corresponding integration, by using
 \beq\label{inty+}
 \int \frac{\rmd \acute y ^+}{[-2\acute y ^+(\dacute y ^- - y^-)+
  1+\bm{y}_\perp^2+i\epsilon]^4}
  \,= \,
  \frac{-\pi i}{3(1+\bm{y}_\perp^2)^3} \,\delta (\dacute y ^- - y^-)\,.
  \eeq
We thus find
 \beq
  \langle \hat{\mathcal E}(y^-,\bm{y}_\perp)\rangle \,\equiv\,
\frac{\sqrt{2}(1+\bm{y}^2_\perp)^3}{8L}\, \langle \hat
T_{--}(y^-,\bm{y}_\perp)\rangle =\,\frac{E_0}{4\pi}\, P(y^-)\,.
  \label{delta}
  \eeq
The new quantity $\langle \hat{\mathcal E}(y^-,\bm{y}_\perp)\rangle$ is
defined in such a way that its integral over $y^-$ yields the energy
density per unit angle, cf. \eqnum{flat}. By integrating the last
expression above over $y^-$ and using the probability conservation
\eqref{prob1}, one recovers the previous result ${\rmd E}/{\rmd \Omega} =
E_0/4\pi$. This meets our expectations: the effects of the longitudinal
string fluctuations are washed out by the integral over $y^-$.

But the crucial new feature in \eqnum{delta} is its explicit dependence
upon $y^-$. Note the remarkable fact that the integral over $\acute y ^+$
in \eqnum{inty+} has identified the longitudinal coordinate of a point on
the string ($\dacute y ^-$) with that of a point on the boundary ($y^-$).
In the absence of fluctuations, $P(y^-)\to \delta(y^-)$, and the above
expression yields, as expected, the shock--wave result in
\eqnum{hattbound} --- that is, a spherical shell of zero width ($t=r$) in
the original $x$ coordinates. But for generic fluctuations, \eqnum{delta}
shows the interesting result that the width of the probability
distribution $P(y^-)$ for the fluctuations acts at the same time as the
width of the distribution in $y^-$ of the energy density on the boundary.
This is the announced relation between fluctuations and broadening.

To better appreciate the physical meaning of this relation, it is useful
to recall that the relation between the coordinates $y^M$ and the
original coordinates $x^M$ looks {\em different} in the bulk ($z>0$) as
compared to the boundary ($z=0$). The argument $y^-$ of $P(y^-)$ refers
to a point on the closed string in the bulk. On the other hand, the
argument $y^-$ of $\langle\hat T_{--}\rangle $ is a boundary variable,
for which $z=0$ and hence
 \beq y^- = \frac{2x^+x^- - \bm{x}_\perp^2}{2x^+L}\,=\,
\frac{t-r}{\sqrt{2} L}\label{i3}
 \qquad\mbox{for}\qquad\bxT=0\,,\eeq
where in the second equality we have used the underlying spherical
symmetry to fix $\bxT=0$ (and hence $r=x^3$). Hence, via \eqnum{delta},
the large string fluctuations in $\acute y^-$  are directly mapped onto
fluctuations $\Delta r$ in the localization of the energy distribution
around a central value $r=t$. Specifically, if the probability
distribution $P(y^-)$ for the string fluctuations has some characteristic
width $\Delta \acute y ^-$, then \eqref{delta} and \eqref{i3} imply that
there will be a width
\beq
\Delta r=L\Delta \acute y ^-\,, \label{spr}
\eeq
for the radial ``energy shell" on the boundary at time $t$.

This broadening of the energy distribution is in agreement with the UV/IR
correspondence, as we now explain. To that aim, we return to the integral
representation (\ref{limit}) but, instead of performing the integral over
$\acute y ^+$ as in \eqnum{inty+}, we consider first the integral over
$\dacute y ^-$. If the probability distribution $P(\dacute y^-)$ has
support in a range $\Delta \acute y ^-$, then the integral over $\dacute
y ^-$ within that range requires $|\acute y^+|\lesssim  1/\Delta \acute
y^-$. But from Eq.~(\ref{cornalba}), $-2\acute y^+=L/\acute x^+$ so that
$\acute x^+ \sim L\Delta \acute y^-$. For the falling particle $\acute
x^+=z/\sqrt{2}$, so we finally get the following estimate
   \beq
   z\sim L \Delta \acute y^- \,,\label{ir}
   \eeq
for the typical values of $z$ contributing to fluctuations with amplitude
$\Delta \acute y^-$. By comparing with \eqnum{spr}, we see that $\Delta
r\sim z$, which is the expected correlation according to the UV/IR
correspondence. Note also that this correlation is washed out
--- once again, as expected
--- when integrating the energy distribution over $y^-$. Indeed, if one
integrates Eq.~(\ref{limit}) over $y^-$, by using a formula analogous to
\eqnum{inty+}, one generates the $\delta$--function $\delta(\acute y
^+)$, which then implies $\acute x^+ =z/\sqrt{2}\to \infty$. Thus, the
{\em angular} energy distribution, as obtained after integration the
radial distribution over $r$ (or $t$), appears to be generated via
backreaction at $z\to\infty$, as already noted  in
Ref.~\cite{Hofman:2008ar}.

Finally, how large is the width $\Delta \acute y ^-$ of the probability
distribution $P(y^-)$ for the string fluctuations ? This can be
identified with the squared root of $\langle (\delta y^-)^2 \rangle$,
which is the equal--point limit of the respective 2--point function. In
flat space at least, $\langle (\delta y^-)^2 \rangle$ shows a strong UV
divergence, so the above question makes no sense without a cutoff on the
number of contributing string modes. Similar, and even stronger, UV
problems are also inherent in the evaluation of $P(\dacute y ^-)$
according to \eqnum{Pdef} : by exponentiating the $\delta$--function in
the integrand there and then expanding the exponential, one generates an
infinite series of higher--point correlations of $ \acute y ^-$, which
are evaluated at coincident points. To obtain an heuristic estimate for
$\Delta \acute y ^-$, let us introduce an upper limit $N$ on the mode
number $n$ for the string oscillations. Requiring $n\le N$, one finds
(see Appendix \ref{SLC})
\beq
\langle(\Delta y^-)^2\rangle \sim \frac{N^2}{(Lp^+)^2}\,, \label{28}
\eeq
which shows a much stronger cutoff dependence than the respective
transverse fluctuations 
\beq
\langle(\Delta y^i)^2\rangle \sim \frac{\ln N}{\sqrt{\lambda}} \,.
\label{27}
 \eeq
If, as argued in \cite{Polchinski:2001ju}, the warping in the 5th
dimension acts effectively as a UV cutoff in AdS$_5$, then it seems
natural to take $N$  of the order
\beq
N\sim e^{\sqrt{\lambda}}\,, \label{29}
\eeq
since via \eqnum{27} this yields a result for $\langle(\Delta
y^i)^2\rangle$ which is finite and independent of $\lambda$, in agreement
with \cite{Polchinski:2001ju}. With this choice for $N$, Eqs.~\eqref{28}
and (\ref{spr}) yield a very large spread in $r$ away from $r=t$.

Of course, in reality we do not have a good control over string
fluctuations when they become so large as shown in
(\ref{28})--(\ref{29}). However, we do believe that we have given good
arguments for the broadening of the radial profile of the outgoing energy
produced by the decay of a time--like photon. Moreover, it appears that
the scale $\Delta \acute y^-$ which controls this broadening according to
(\ref{spr}) is large and not suppressed by inverse powers of $\lambda$.
This implies strong modifications of the predictions of the supergravity
approximation, whose accurate calculation would require a proper
treatment of string fluctuations in AdS$_5$.

\section*{Acknowledgments}

During the late stages of this work, three of us (E.~I., A.H.~M., and
D.N.~T.) have benefited from financial support from the {\em Erwin
Schr{\" o}dinger International Institute for Mathematical Physics} in
Vienna, within the context of the {\sf ESI Programme on AdS Holography
and the Quark-Gluon Plasma} (Vienna, 2 August--29 October 2010),
organized by A.~Rebhan, S.~Husa, and K.~Landsteiner. We would like to
thank ESI and the organizers of the Programme for their hospitality and
for offering us most suitable conditions for work and for intense and
fruitful discussions. In particular, during our stay in Vienna we have
enjoyed and benefited from discussions with J.~Casalderrey--Solana,
P.~Chesler, E.~Kiritsis, I.~Klebanov, R.~Myers, D.~Teaney, and A.~Yarom.
We thank J.~Casalderrey--Solana, G.~Giecold and U.~G{\" u}rsoy for useful
comments on the manuscript. The work of Y.~H. is supported by Special
Coordination Funds for Promoting Science and Technology of the Ministry
of Education, the Japanese Government. The work of E.~I. is supported in
part by Agence Nationale de la Recherche via the programme
ANR-06-BLAN-0285-01. The work of A.H.~M. is supported in part by the US
Department of Energy.

\appendix
\section{The constant velocity string: a different perspective}
 \label{appheavy}

In this Appendix we present an alternative way to derive the energy
energy density induced by a heavy quark moving with constant velocity.
Considering Eqs.~\eqref{EAunif} and \eqref{Wunif}, we make the change of
variables
 \beq\label{ttqz}
  \tp = t_q +\gamma z \quad\Rightarrow\quad \dif \tp = \dif t_q, \eeq
and then $\mcal{W}$ becomes a linear function of $z$, more precisely
 \beq
\mcal{W} = \mcal{W}_q + 2 \gamma z \left(t - \upsilon x -
\frac{t_q}{\gamma^2}\right), \eeq where we have defined \beq \mcal{W}_q =
-(t- t_q)^2 + (x- \upsilon t_q)^2 + \bxT^2.
\eeq
Now it is straightforward to integrate over $z$ making use of the
$\delta$-function which will set
\beq\label{zWq} z = -\frac{\mcal{W}_q}{2 \gamma (t - \upsilon x
-t_q/\gamma^2)}.
 \eeq
Causality requires that the denominator in the above \eqnum{zWq} should
be positive and, since $z \geq 0$, we need $W_q \leq 0$ in order to
obtain a non-zero result. Thus we have
 \beq \EA = \frac{\sqrt{\lambda}}{2
\pi^2}\, \gamma^2 (3 - \upsilon^2) \del^2_{\bxT^2}
\int_{-\infty}^{\infty} \dif t_q \frac{\Theta(- \mcal{W}_q)}{\gamma^2 (t
- \upsilon x) - t_q}.
 \eeq
It is clear that the first derivative will act on the step function, that
is, on the upper end point of the integration interval, to give
 \beq
\del_{\bxT^2} \Theta(-\mcal{W}_q) = - \delta(\mcal{W}_q) = - \frac{\gamma
\delta[t_q - \gamma^2(t - \upsilon x) + \gamma \sqrt{\bxT^2 + \gamma^2 (x
 - \upsilon t)^2}]} {2 \sqrt{\bxT^2 + \gamma^2 (x - \upsilon t)^2}}. \eeq
Now also the integration over $t_q$ can be easily done by making use of
the $\delta$-function in the above. We find
 \beq \EA =
-\frac{\sqrt{\lambda}}{4 \pi^2}\, \gamma^2 (3 - \upsilon^2) \del_{\bxT^2}
\frac{1}{\bxT^2 + \gamma^2 (x - \upsilon t)^2},
 \eeq
which leads to \eqnum{EAHQ1} in the main text, and similarly we can
calculate $\EB$ which leads to \eqnum{EBHQ}. What is important to notice
in the above derivation, is that the $t_q$-integration is determined by
$\mcal{W}_q=0$, which in turn fixes $z=0$ when performing the
$z$-integration.

\section{Boundary to bulk scalar propagator in AdS$_5$}
\label{appProp}

Here we would like to construct the boundary to bulk scalar propagator.
We shall obtain it from the $-\zp^4/4$ coefficient of the bulk to bulk
propagator. In momentum space, and for space--like kinematics, the latter
is given by (see for example Appendix B in \cite{Avsar:2009xf})
 \beq
 G(\omega,\bq,z,\zp) = -z_<^2\, z_>^2 \rmI_2(Qz_<) \rmK_2(Qz_>),
 \eeq
where $Q^2 = \bq^2 - \omega^2>0$. In the time--like region $Q^2<0$ the
propagator is obtained by analytic continuation together with the
retardation prescription $\omega \to \omega + \rmi 0$. Then we find for
the boundary to bulk propagator
 \beq\label{Dmom}
 D(\omega,\bq,z) =
 \begin{cases}
 \displaystyle{\frac{z^2 Q^2}{2}\,\rmK_2(Q z)}
 &\quad \text{if} \quad Q^2>0,
 \\*[0.25cm]
 \displaystyle{\frac{\rmi \pi z^2 |Q|^2}{4}\,\rmH^{(1)}_2(|Q| z)}
 &\quad \text{if} \quad Q^2<0,\,\omega>0,
 \\*[0.25cm]
 -\displaystyle{\frac{\rmi \pi z^2 |Q|^2}{4}\,\rmH^{(2)}_2(|Q| z)}
 &\quad \text{if} \quad Q^2<0,\, \omega<0.
 \end{cases}
 \eeq
Notice that $D(\omega,\bq,z=0)=1$. Now we wish to Fourier transform back
to configuration space. Performing the angular integrations in spherical
coordinates we find
 \beq
 D(\omega,\br,z) =
 \int \frac{\dif^3\bq}{(2\pi)^3}\,\rme^{\rmi \bq \cdot \br} D(\omega,\bq,z)
 = \frac{1}{2\pi^2 r}
 \int_{0}^{\infty}
 \dif q\,q \sin(q r) D(\omega,\bq,z)
 \eeq
with $r=|\br|$ and $q=|\bq|$. For $\omega>0$ we can use the first two
cases in \eqnum{Dmom} and integrating over $q$ we obtain
 \beq\label{Domega}
 D(\omega>0,\br,z) = \frac{\rmi}{8 \sqrt{2\pi}}\,\frac{z^4}{\rho^{7/2}}\,\omega^{7/2}\,\rmH_{7/2}^{(1)}(\omega \rho),
 \eeq
where we have defined $\rho = \sqrt{r^2 + z^2}$. We notice that
$\rmH_{7/2}$ is a rather simple function which for $\omega>0$ is given by
 \beq\label{H72}
 \rmH_{7/2}^{(1)}(\omega\rho) = \sqrt{\frac{2}{\pi\omega\rho}}\,\rme^{\rmi \omega \rho}
 \left[
 1 - \frac{6}{\rmi \omega \rho} + \frac{15}{(\rmi \omega\rho)^2} -\frac{15}{(\rmi\omega\rho)^3}
 \right].
 \eeq
For $\omega<0$ we have
 \beq
 D(\omega<0,\br,z) = D^*(|\omega|,\br,z),
 \eeq
and therefore we can write
 \beq
 D(t,\br,z) =
 2 \int \frac{\dif \omega}{2\pi}\,{\rm Re} \left[ \rme^{-\rmi \omega t} D(\omega,\br,z)\right]
  \eeq
Making use of Eqs.~\eqref{Domega} and \eqref{H72} we see that the
integrand in the above is an even function of $\omega$ and therefore we
can multiply by a factor of $1/2$ and extend the integration to the whole
real axis. Furthermore the terms odd in $\omega$ will vanish
automatically and the result will be real. Thus
 \beq
 D(t,\br,z) = -\frac{z^4}{8 \pi \rho^4}
 \int_{-\infty}^{\infty} \frac{\dif \omega}{2\pi}\,(\rmi\omega)^3
 \rme^{\rmi \omega (\rho - t)}
 \left[
 1 - \frac{6}{\rmi \omega \rho} + \frac{15}{(\rmi \omega\rho)^2} -\frac{15}{(\rmi\omega\rho)^3}
 \right].
 \eeq
The last expression is simply related to an integral representation of
the $\delta$-function and its derivatives and we have
 \beq
 D(t,\br,z) = -\frac{z^4}{8 \pi \rho^4}
 \left[ \delta'''(\rho - t) -
 \frac{6\delta''(\rho - t)}{\rho}
 +\frac{15\delta'(\rho - t)}{\rho^2} -
 \frac{15\delta(\rho - t)}{\rho^3}
 \right].
 \eeq
It is a straightforward exercise to start from $16 \rho^4 \Theta(t)
\delta'''(\rho^2 - t^2)$ and show that it is equal to the square bracket
above (in the sense of distributions). Thus we finally have
 \beq\label{Dconfig}
 D(t,\br,z) = -\frac{2 z^4}{\pi}\,
 \Theta(t)\, \delta'''(\rho^2 - t^2).
 \eeq

\section{Backreaction in the non--relativistic limit}
\label{appPulse}

Let us add here some complemental details in the calculation to the
energy radiated by a non--relativistic quark when it goes under an
arbitrary 1D motion. The string embedding functions and their derivatives
are given by
 \beq
 X^M = (t,x_s,0,0,z),
 \qquad \dot{X}^M = (1,\dot{x_s},0,0,0),
 \qquad {X^M}' = (0,{x'_s},0,0,1),
 \eeq
and we easily find the components and the determinant of the induced
metric on the world--sheet (with the usual parametrization
$(\tau,\sigma)\equiv (t,z)$)
 \beq
 g_{\tau\tau} &=& \dot{X} \cdot \dot{X} = |G_{00}|\,(-1 + \dot{x}_s^2),
 \\*[0.15cm]
 g_{\sigma\sigma} &=& X' \cdot X' = |G_{00}|\,(1 + x_s^{\prime 2}),
 \\*[0.15cm]
 g_{\tau\sigma} &=& \dot{X} \cdot X' = |G_{00}|\, \dot{x}_s x'_s,
 \\*[0.15cm]
 \sqrt{-g} &=& |G_{00}|\, \sqrt{1-\dot{x}_s^2+x_s^{\prime 2}}.
 \eeq
Now starting from \eqnum{tMN} we can find
 \beq
 \tilde{t}^{MN} = \frac{T_0}{\sqrt{-g} \sqrt{-G}}\,
 \left[
 g_{\sigma\sigma} \dot{X}^M \dot{X}^N +
 g_{\tau \tau} {X^M}' {X^N}' -
 g_{\tau \sigma} \left(\dot{X}^M {X^N}' + {X^M}' \dot{X}^N\right)
 \right].
 \eeq
Let us calculate the components of the above expression and at the same
time lower the indices. Since the metric is diagonal we do this by
multiplying each component by $\pm |G_{00}|^2$, where we use the minus
only when one of the two indices is equal to 0. After substitution of the
common coefficient
 \beq
 \frac{T_0 |G_{00}|^3}{\sqrt{-g}\sqrt{-G}} =
 \frac{\sqrt{\lambda}}{2 \pi}\, \frac{z}{L^3}\, \frac{1}{\sqrt{1-\dot{x}_s^2+x_s^{\prime 2}}},
 \eeq
we find
 \beq\label{tmn string}
 \tilde{t}_{00} &=& \frac{\sqrt{\lambda}}{2 \pi}\, \frac{z}{L^3} \,\frac{1 + x_s^{\prime 2}}{\sqrt{1-\dot{x}_s^2+x_s^{\prime 2}}},
 \qquad
 \tilde{t}_{01} = -\frac{\sqrt{\lambda}}{2 \pi}\, \frac{z}{L^3}\, \frac{\dot{x}_s}{\sqrt{1-\dot{x}_s^2+x_s^{\prime 2}}},
 \nn
 \tilde{t}_{05} &=&
 \frac{\sqrt{\lambda}}{2 \pi}\, \frac{z}{L^3} \,
 \frac{\dot{x}_s x_s^{\prime}}{\sqrt{1-\dot{x}_s^2+x_s^{\prime 2}}},
 \qquad
 \tilde{t}_{11} = \frac{\sqrt{\lambda}}{2 \pi}\, \frac{z}{L^3} \,\frac{\dot{x}_s^2- x_s^{\prime 2}}{\sqrt{1-\dot{x}_s^2+x_s^{\prime 2}}},
 \nn
 \tilde{t}_{15} &=& -\frac{\sqrt{\lambda}}{2 \pi}\, \frac{z}{L^3} \,\frac{x_s^{\prime}}{\sqrt{1-\dot{x}_s^2+x_s^{\prime 2}}},
 \qquad\!\!\!\!
 \tilde{t}_{55} = \frac{\sqrt{\lambda}}{2 \pi}\, \frac{z}{L^3}\,\frac{-1 + \dot{x}_s^2}{\sqrt{1-\dot{x}_s^2+x_s^{\prime 2}}}.
 \eeq
Expanding the square root and keeping only the quadratic terms of the
components we have
 \beq
 \tilde{t}_{00} = \tilde{t}_{55}= \frac{\sqrt{\lambda}}{4 \pi}\,\frac{z}{L^3}\,
 (\dot{x}_s^2 + x_s^{\prime 2}),
 \quad
 \tilde{t}_{05} = \frac{\sqrt{\lambda}}{2 \pi}\,\frac{z}{L^3}\,
 \dot{x}_s x_s^{\prime},
 \quad
 \tilde{t}_{11} = \frac{\sqrt{\lambda}}{2 \pi}\,\frac{z}{L^3}\, (\dot{x}_s^2 - x_s^{\prime 2}),
 \eeq
and using \eqnum{xs der} to express $\dot{x}_s$ and $x'_s$ in terms of
the boundary motion we finally arrive at \eqnum{tradPulse}. We should
notice here that $\tilde{t}_{01}$ and $\tilde{t}_{15}$ contain a linear
term in $\dot{x}_s$ and $x'_s$ respectively. However, they do not contain
a term quadratic in the boundary motion since the first correction to the
solution to the string EOM \eqnum{EOM} is cubic in the boundary motion.
Indeed, one can check by direct substitution that
 \beq\label{xs nonlinear}
 x_s(t,z) = x_q(t-z) + z \dot{x}_q(t-z)
 - \frac{1}{2}\,z^2\,\dot{x}_q^2(t-z) \ddot{x}_q(t-z),
 \eeq
is the solution to the required order of accuracy. But in general, one
might consider the situation where terms in $t_{MN}$ which are constant
or linear in $\dot{x}_s$ and $x'_s$  combine with the expansion of
$\delta(\mcal{W})$ in powers of $x_s$ and therefore produce quadratic
terms which can give rise to radiation (recall that we have set $x_s=0$
in the argument of $\mcal{W}$ in Sect.~\ref{SmallPulse}). However one can
find that
 \beq
 \delta(\mcal{W}) = \frac{1}{2 \rho}\, \delta(\tp - t +\rho)
 + \frac{x x_s}{2 \rho^3} \left[\delta(\tp - t +\rho) - \rho\,\delta'(\tp - t +\rho)\right] + \order{x_s^2},
 \eeq
with $\rho = \sqrt{z^2 + r^2}$, and such terms will be eventually
suppressed by inverse power of $r$ and thus do not contribute to the
radiated energy.

\section{Strings in the light-cone gauge}
\label{SLC}

Here we remind the reader how the free string in flat space is quantized
and evaluate the size of the fluctuations (for more details see e.g.
\cite{Polchinski:1998rq,Zwiebach:2004tj}). As usual, we denote the
world--sheet coordinates by $\tau$ and $\sigma$, with $0\le\sigma\le
2\pi$ for a closed string. We shall work in the light--cone (LC) gauge,
defined by
\beq
y^+=\frac{\alpha'p^+\tau}{L}\,,
\eeq
($p^+$ denotes the LC momentum of the string) together with the
associated $\sigma$ parametrization. In this gauge, the transverse
coordinates $\bm{y}_\perp(\tau,\sigma)=(y^1,y^2)$ of the string obey the
equations of motion of harmonic oscillators,
 \beq\label{harmonic}
 \left(\frac{\del^2}{\del\tau^2}\,-\,\frac{\del^2}{\del\sigma^2}\right)
 y^i(\tau,\sigma)\,=\,0\,,\eeq
whereas the longitudinal coordinate $y^-$ is not independent, but rather
it is related to the transverse coordinates by the constraints enforcing
the LC gauge. Namely, one has (for a closed string)
\beq\label{constraint}
\partial_\tau y^- = \frac{L}{2\alpha' p^+}
\left( (\partial_\tau \bm{y}_\perp)^2+
(\partial_\sigma \bm{y}_\perp)^2\right)\,.
\eeq
\eqnum{harmonic} is solved as (below, $i=1,\,2$)
\beq
y^i(\tau,\sigma) = y^i_0 + \sqrt{\frac{\alpha'}{2L^2}}(\alpha_0^i +
\tilde{\alpha}_0^i)\tau + \rmi\sqrt{\frac{\alpha'}{2L^2}} \sum_{n\neq 0}
\left\{ \frac{\alpha_n^i}{n} \,\rme^{-\rmi n(\sigma + \tau)} +
\frac{\tilde{\alpha}_n^i}{n}\, \rme^{\rmi n(\sigma-\tau)} \right\}\,.
\eeq
The $\alpha$ operators with/without a tilde refer to closed string waves
which are left/right moving along the string. The quantization of the
transverse string fluctuations proceeds in the standard way, by imposing
 \beq [\alpha_m^i, \alpha_n^j]=m\delta^{ij}\delta_{m,-n}\,,
 \eeq
together with the similar relation for the tilded operators. (Tilded and
untilded operators commute with each other.) Notice that the standard
creation and annihilation operators for the quantum harmonic oscillator
are related to the $\alpha_n^i$ operators above via
 \beq
 a^i_n\,=\,\frac{\alpha_n^i}{\sqrt{n}}\,,\qquad
 a^{i\dagger}_n\,=\,\frac{\alpha_{-n}^i}{\sqrt{n}}\,,\quad
 \mbox{with}\quad n\ge 1\,,\eeq
and obey indeed $[a^i_n,a^{i\dagger}_m]=\delta^{ij}\delta_{mn}$.

The constraint equation \eqref{constraint} can be integrated to yield
\beq
y^-(\tau,\sigma)=y^-_0 + \frac{\tau}{2p^+L}(L_0^\perp +
\tilde{L}_0^\perp) +\frac{\rmi}{2p^+L}\sum_{n\neq 0}
\frac{1}{n}\left(L_n^\perp\, \rme^{-\rmi n(\sigma+\tau)} +
\tilde{L}_n^\perp\, \rme^{\rmi n(\sigma-\tau)} \right)\,,
\eeq
 where
\beq
L_n^\perp = \frac{1}{2}\sum_p \alpha_{n-p}^i \alpha_p^i\,, \qquad
L^\perp_{-n}=(L_n^\perp)^\dagger\,,
\eeq
are the so-called transverse Virasoro generators which satisfy the
commutation relation
\beq
[L_m, L_n] = (m-n)L_{m+n} + \frac{c}{12}(m^3-m)\delta_{m,-n}\,.
\eeq
$c$ is the central charge which equals the number of the transverse
dimensions $D-2$.

The two--point functions of these operators are evaluated in the standard
way:
\beq\label{yi2pt}
\langle y^i(\tau,\sigma),y^j(0,0)\rangle=
\delta^{ij}\frac{\alpha'}{2L^2}\sum_{n=1}^\infty \frac{1}{n}
\left(\rme^{-in(\sigma + \tau)} + \rme^{in(\sigma-\tau)} \right)\,,
\eeq
\beq\label{y-2pt}
 \langle y^-(\tau,\sigma)y^-(0,0)\rangle &=& \frac{c}{48(p^+L)^2}
\sum_{n=1}^\infty \frac{n^2-1}{n}\left(\rme^{-in(\sigma+\tau)}
 + \rme^{in(\sigma-\tau)}\right)\,.
\eeq
From \eqnum{yi2pt} it is clear that the size of the transverse
fluctuation $\delta y^i$ due to a single mode $n$ is
 \beq\label{dyi}
 \delta y^i_n\,\sim\,\sqrt{\frac{\alpha'}{nL^2}}\,=\,
 \frac{1}{\sqrt{n\sqrt{\lambda}}}\,,\eeq
which is very small when $\lambda$ is large. On the other hand,
\eqnum{y-2pt} implies that a single mode gives a longitudinal fluctuation
with typical size
 \beq\label{dy-}
 \delta y^-_n\,\sim\,\frac{1}{p^+ L}\,,\eeq
which is {\em not} suppressed when $\lambda$ is large and which gets much
larger contributions from large $n$ as compared to $\delta y^i$. This
last feature becomes especially important whenever one needs to consider
the equal--point limit of the 2--point function (or, more generally, of a
$n$--point function), so like in the discussion of the average LC energy
$\langle \hat{\mathcal E}\rangle$ in Sect.~\ref{Longit}.

Specifically, when taking the limit $\tau,\sigma\to 0$ in
Eqs.~\eqref{yi2pt} and \eqref{y-2pt} one encounters ultraviolet
divergences coming from the sum over the large--$n$ modes, which are {\em
logarithmic} in the case of \eqnum{yi2pt}, but {\em quadratic} in the
case of \eqnum{y-2pt}. Introducing a mode cutoff $n\le N$ to regularize
the divergence, we get
\beq
\langle (y^i)^2\rangle = \frac{\alpha'}{L^2}\ln N = \frac{\ln N}{
\sqrt{\lambda}} \,, \label{suss}
\eeq
\beq
\langle (y^-)^2 \rangle = \frac{cN^2}{48(p^+L)^2}\,. \label{fluct}
\eeq
What should be the value of the mode cutoff $N$ ? If we were to consider
a scattering problem --- the scattering between two strings --- then,
first, longitudinal modes would not matter and, second, the effective
region in $\tau$ would be non--zero and fixed by the kinematics:
$\delta\tau \sim 1/(\alpha' s)$ with $\sqrt{s}$ the center of mass energy
of the scattering. This limits the participating modes to $n\le N=\alpha'
s$ and the corresponding fluctuation sizes to
 \beq\label{suss1}
\langle (y^i)^2\rangle \,\sim\, \frac{\ln \alpha' s}{ \sqrt{\lambda}}\,,
\qquad \langle (y^-)^2 \rangle \,\sim\, \frac{(\alpha' s)^2}{(p^+L)^2}\,.
\eeq
The first equation in \eqref{suss1} gives the shrinkage of the
diffraction peak while the second one gives the natural longitudinal
extent of the scattering process.

However, when evaluating $\langle \hat{\mathcal E}\rangle$ in \eqnum{ali}
or \eqref{limit}, one encounters higher powers of  $y^-(\tau,\sigma)$
which must be averaged and there is no natural cutoff. Polchinski and
Susskind \cite{Polchinski:2001ju} argue that in AdS$_5$ the string
fluctuations are cutoff by warping in the fifth dimension, in such a way
that (\ref{suss}) becomes finite and of order one (and thus independent
of $\sqrt{\lambda}$). This suggests
\beq
N\sim \rme^{\sqrt{\lambda}}\,,
\eeq
as a cutoff on modes which in turn would give, cf. \eqnum{fluct},
 \beq
 \Delta y^- \sim \frac{e^{\sqrt{\lambda}}}{p^+L}\,,
 \eeq
which is a very large value. Note that
 \beq
 x\sim e^{-\sqrt{\lambda}}\,,
 \eeq
is the region in Bjorken--$x$ variable where one starts to see ``partons"
in DIS at strong coupling \cite{HIM1}. Then
 \beq
 \Delta y^- \sim \frac{1}{xp^+ }\,,
 \eeq
can be identified with the lifetime of partons which have energy $xp^+$
-- here the partons which compose the closed string, or the `falling
point particle'.

\providecommand{\href}[2]{#2}\begingroup\raggedright\endgroup


\end{document}